\documentclass[aps,10pt,twocolumn,superscriptaddress,preprintnumbers,amsmath,amssymb,prl]{revtex4-1}

\usepackage{times}
\usepackage[pdftex]{graphicx}
\usepackage[bookmarksopen=true,bookmarksopenlevel=1]{hyperref}
\usepackage{epstopdf}
\usepackage{amsmath}
\usepackage{amsfonts}
\usepackage{amssymb}
\usepackage{graphicx}
\usepackage{sidecap}
\sidecaptionvpos{figure}{c}
\usepackage{color}
\usepackage{dcolumn}
\usepackage{bm}
\usepackage[compact]{titlesec}
\usepackage[normalem]{ulem}
\usepackage{bookmark}
\usepackage{natbib}
\usepackage{indentfirst}
\usepackage{appendix}

\definecolor{mypink}{RGB}{219, 48, 122}
\definecolor{mygreen}{RGB}{51, 153, 102}
\definecolor{brown}{RGB}{165, 42, 42}
\def\braun{\color{black}}
\def\braunb{\color{black}}

\newlength{\beforesection}
\newlength{\aftersection}
\newlength{\beforesubsection}
\newlength{\aftersubsection}
\titlespacing*{\section}{0pt}{\beforesection}{\aftersection}
\titlespacing*{\subsection}{0pt}{\beforesubsection}{\aftersubsection}

\def\dirGM{$\overline{\rm \Gamma}-\overline{\rm M}$}

\def\root33{$\sqrt{3}\times\sqrt{3}$ {\it R}30$^\circ$}
\def\RT3{$\sqrt{3}$}
\def\BiSbTe{(Bi,Sb)$_2$Te$_3$}
\def\BiTe{Bi$_2$Te$_3$}
\def\BiSe{Bi$_2$Se$_3$}
\def\Gbar{$\overline{\rm \Gamma}$} 
\def\Xbar{$\overline{\rm X}$}


\begin{document}

\title{Angle-resolved photoemission of topological materials}

\author{Jaime S\'anchez-Barriga}
\affiliation{Helmholtz-Zentrum Berlin f\"ur Materialien und Energie, Albert-Einstein-Str.~15, 12489 Berlin, Germany}
\affiliation{IMDEA Nanoscience, C/ Faraday 9, Campus de Cantoblanco, 28049, Madrid, Spain}

\author{Oliver J. Clark}
\affiliation{Helmholtz-Zentrum Berlin f\"ur Materialien und Energie, Albert-Einstein-Str.~15, 12489 Berlin, Germany}

\author{Oliver Rader}
\affiliation{Helmholtz-Zentrum Berlin f\"ur Materialien und Energie, Albert-Einstein-Str.~15, 12489 Berlin, Germany}

\begin{abstract} 
Topological materials have gained significant attention in condensed matter physics due to their unique electronic and transport properties. Three-dimensional (3D) topological materials are characterized by robust electronic states that are protected by symmetries and exhibit peculiar spin textures. They offer a rich platform for for future information technology including spintronics and topological quantum computing.  Here, we review the investigation by angle-resolved photoelectron spectroscopy (ARPES) of topological phases such as 
strong topological insulators,   topological crystalline insulators, magnetic topological insulators, and 3D Dirac, Weyl, nodal, {\braun and chiral} semimetals {\braunb and address the status of correlated topological insulators and topological superconductors.} {\braun  A special emphasis is laid on examples from the transition metal dichalcogenide family.}
{\braunb Moreover, insights from ultrafast pump-probe experiments are reviewed and a brief outlook is provided.}
\end{abstract}

\maketitle


\begin{itemize}
  \item   ARPES plays a pivotal role as the primary experimental method for investigating the electronic structure and topology of 3D topological materials.
 \item    ARPES reveals topologically protected surface states as well as their origin in the bulk band structure due to its 
 surface and bulk sensitivity, wave vector resolution in 3D, and sensitivity to symmetries.
 \item    Spin resolution and circular dichroism in ARPES offer valuable information about degeneracies, spin-momentum locking, and spin and orbital textures but  final-state effects need to be considered.
 \item    ARPES enables conclusions on fundamental aspects of topological materials, such as band inversion, 
spin-orbit interaction, and the Berry phase.
      
 \item      The use of ARPES provides  insights into exchange splittings, superconducting pairing symmetries, and the robustness of topological surface states, including surface band bending,  interface properties, and quantization effects.

\item ARPES has unambiguously identified examples for several classes of topological materials with 
correlated topological insulators and topological superconductors still outstanding.

\item  Transition metal dichalcogenides  introduce additional flexibility in the study of topological materials and offer unique opportunities for exploration.

 \item    ARPES allows conclusions on scattering channels and aids in understanding the behavior of spin and charge carriers in topological matials. 

\item  The application of ultrafast pump-probe ARPES allows for the investigation of unoccupied states, their spin texture, and the dynamics of interactions between bulk and surface states, especially under external stimuli like infrared light. It can be utilized to study the creation of charge and spin-polarized currents as well as the emergence of exotic photon-dressed states in topological materials.

 \item    The development of micro- and nano-ARPES techniques opens up possibilities for studying smallest crystallites and domains, while additional in-operando options allow to study the effects of current, voltage, and strain and will enable the exploration of topological phase diagrams comprehensively. 
  
\end{itemize}


\section{{\normalfont\normalsize\bfseries I. INTRODUCTION}}

Topological phases of matter are an important research area in condensed matter physics due to their unique electronic and transport properties. 
In direct analogy the broader mathematical concept describing objects more generally, the topology of a system's electronic structure remains unchanged unless the system is driven through a topological phase transition. A change in the number of holes in the object corresponds then to a change in parameters such as the quantity of gapless surface states in the electronic structure. As such, the  electronic states characterizing a non-trivial topological phase are extremely robust and insensitive to smooth changes in material parameters. Most typically, in 3D (2D) systems, these states manifest as highly conductive Dirac cone-like surface (edge) states with unconventional spin textures. Indeed, 2D topological phases such as  the quantum Hall, the quantum spin Hall, and the quantum anomalous Hall insulator, have been discovered and investigated through their quantized lossless 1D edge channel transport. Similarly, 3D topological phases have mainly been identified by angle-resolved photoelectron spectroscopy (ARPES) studies into the associated topologicaly projected surface states spanning across the topologically non-trivial band gap, or between so called `Weyl nodes' which serve as sources of topological charge.

ARPES provides energy and momentum-resolved information about the occupied electronic structure, allowing for the detection of the electronic states that are characteristic of topological insulators (TIs) and other topologically non-trivial systems. The method is highly surface-sensitive, thus providing direct access to the protected surface states in addition to the bulk band structure, which are connected through the bulk-boundary correspondence (i.e. the symmetry-mandated presence of surface or edge states at the interface between the non-trivial and trivial media). The 3D momentum resolution enables direct observations regarding dimensionality, and ARPES can be combined with spin detection techniques resolve characteristic spin textures. Moreover, ARPES can be used in time-resolved experiments, allowing for the investigation of the dynamics of topological electronic states under different conditions, such as in response to external perturbations.  

While ARPES was crucial in the characterisation of the early 3D TIs, further topological classifications have since been populated with experimental evidence of materials exhibiting non-trivial band topology beyond that of the strong TIs,
 including topological crystalline insulators, 3D Dirac semimetals, and Weyl semimetals and several other  topological systems, each possessing unique electronic properties and unique spin textures.  Together, topologically non-trivial materials are at the forefront of the research and development of next-generation information technologies with electronic, spintronic and quantum computing applications. 

In the present Chapters, we will address the different types of topological materials in the light of their investigation by ARPES. We will highlight where ARPES can provide insight and which aspects of topological materials facilitated or enabled key ARPES experiments. Likewise, topological materials for which ARPES data are not available are discussed. For an introduction to the ARPES method the reader is referred to excellent reviews such as the one by Comin and Damascelli \cite{CominReview13}. A recent introduction to ARPES with a chapter on topological materials was published by Zhang et al. \cite{ZhangNatRev22} while the review by Lv et al. of ARPES of topological materials has a focus on instrument aspects \cite{LvNatRevPhys19}. ARPES of quantum materials including topological materials has been reviewed by Sobota et al. \cite{SobotaRMP21} and spin-resolved ARPES of topological materials   by Dil \cite{DilReview19}. Topological materials are introduced in other Chapters of the present volume as well as in prominent review articles \cite{HasanKane10,QiZhangRMP11}. Further review articles are pointed out in the individual paragraphs below.

\section{{\normalsize\bfseries II. TOPOLOGICAL INSULATORS}} 

\subsection{{\normalfont\normalsize\bfseries II. A. Prototypical strong topological insulators}}
 
3D TIs behave as a bulk insulator and a surface conductor simultaneously. Their insulating bulk property is due to an inverted band structure between the valence band in the conduction band, typically mediated by the spin-orbit interaction, and their metallic surface is due to gapless surface Dirac cones with spin-momentum locked spin textures, protected by time-reversal symmetry.  This is the situation of strong topological insulators in the $Z_2$ class \cite{HasanKane10,QiZhangRMP11}. 

The bulk-boundary correspondence, responsible for the enforced existence of these topological surface states (TSSs), ensures a coupling between the bulk and surface. This can be exploited in the verification and characterisation of the topological properties of a material. 
The number of band inversions within the bulk corresponds directly to the number of Fermi level crossings of individual {\braun spin sub-bands} 
at the surface. Therefore, the topology of a material can be directly determined from the enumeration of Fermi level crossings along the relevant momentum paths in ARPES data, further cementing the technique as an invaluable tool in the characterisation of ground-state electronic structures. 
 
A strong 3D TI phase was experimentally discovered by ARPES and spin-resolved ARPES in the Bi$_{1-x}$Sb$_{x}$ semiconducting alloy system, demonstrating the existence of non-trivial band topology in a 3D material for the first time \cite{Hsieh08}. The topological classification, {\braunb as subsequently confirmed \cite{Nishide10,Benia15},} was identified by a nontrivial $\mathbb{Z}_2$ topological invariant $\nu_0$=1, describing an odd number of Dirac cones, distinguishing strong $\mathbb{Z}_2$ TIs from trivial or topologically weak materials where the number of Dirac cones is instead even or zero ($\nu_0$=0). 
 
In ARPES, the distinct properties of a 3D topological insulator can be experimentally verified by observing the circular Fermi surface contours around time-reversal invariant momenta or by directly observing their Dirac-like energy-momentum dispersion and spin texture. For a strong $\mathbb{Z}_2$  TI, these signatures must be consistent with an odd number of Fermi-level crossings on the surface, an odd number of band inversions in the bulk, and Dirac-cone TSSs with a chiral, momentum-locked, spin texture that fulfils a Berry phase of $\pi$.  This condition also holds in the cases where the band inversion occurs below the Fermi level, instead requiring the quantification of crossings in the constant energy contour where the inversion takes place. However, the system will not benefit from the conductive surface states in transport. 

\begin{figure*}[t]
	\centering
	\includegraphics[width=\textwidth]{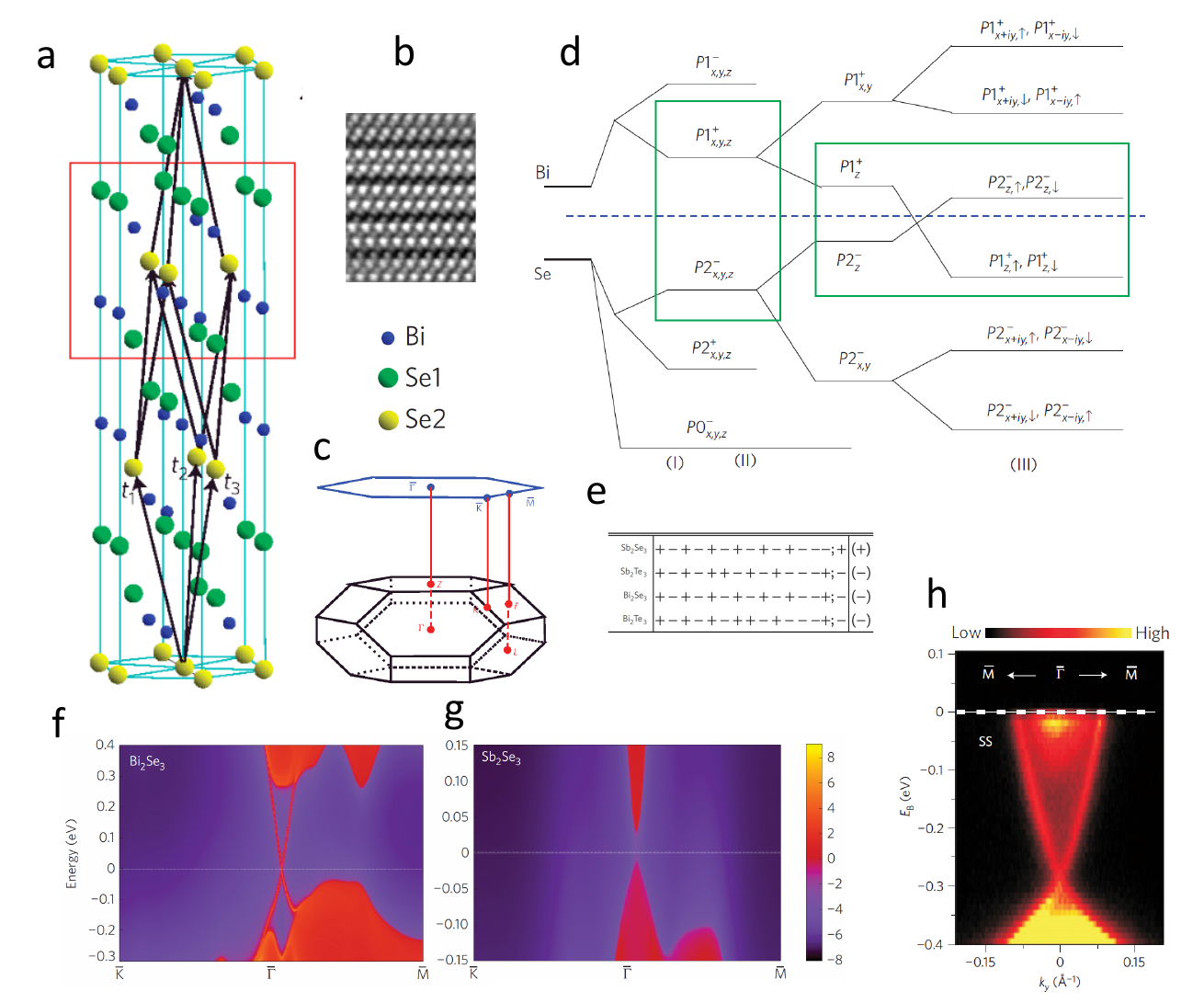}
	\caption{\label{FigBi2Se3} 
Prototypical topological insulator Bi$_2$Se$_3$. (a) The structure consists of quintuple layers (red box) as seen by (b) transmission electron microscopy.  Bi$_2$Se$_3$ shows a single Dirac cone surface state at the center of the (c) surface Brillouin zone. (d) The band inversion occurs due to the spin-orbit interaction included in the rightmost column.  
(e) The parity of the band at the $\Gamma$ point for the four materials Sb$_2$Te$_3$, Sb$_2$Se$_3$, Bi$_2$Se$_3$ and Bi$_2$Te$_3$. The  parities of fourteen occupied s and p bands are shown, and the lowest unoccupied band. The product of the parities for the fourteen occupied bands is given in parentheses on the right of each row. All of these systems, except for Sb$_2$Se$_3$, which has the lowest spin-orbit interaction,  are topological insulators. (f,g) Surface band structure calculations show a Dirac cone surface state for Bi$_2$Se$_3$ and not for Sb$_2$Se$_3$. (h) Observation of the Dirac cone of Bi$_2$Se$_3$ at 
$h\nu=22$ eV. Panels a, c, d, e, f, g from \cite{ZhangNP09}, b (G. Springholz et al., unpublished), h \cite{XiaNP09}. }
\end{figure*}

The most prominent examples of strong  $\mathbb{Z}_2$ TI materials are Bi$_2$Se$_3$, Sb$_2$Te$_3$, and Bi$_2$Te$_3$ which were theoretically predicted \cite{ZhangNP09} and subsequently experimentally verified by ARPES  (Bi$_2$Se$_3$ by \cite{XiaNP09,ChenYLScience09} and Sb$_2$Te$_3$ by \cite{WangG10}) and spin-resolved ARPES \cite{HsiehspinNature09}. An overview of the topologically nontrivial electronic structure of these materials is highlighted in  Fig.~\ref{FigBi2Se3} following their first theoretical prediction. The real space crystal structure of these compounds consists of 2D van der Waals bonded layers of formula units X-M-X-M-X (X=Se,Te\, M=Bi,Sb). Compared with the Bi$_{1-x}$Sb$_{x}$ semiconducting alloy series, these layered stoichiometric compounds possess {\braun a much simpler} electronic structure with only one bulk band inversion, a considerably wide band gap ({\braunb which has been characterized as direct in Bi$_2$Se$_3$ using ARPES \cite{NechaevPRB13}}) and surface states that form a single gapless Dirac cone at the $\overline{\rm\Gamma}$ point of the surface Brillouin zone.  

\begin{figure*}[t]
	\centering
	\includegraphics[width=\textwidth]{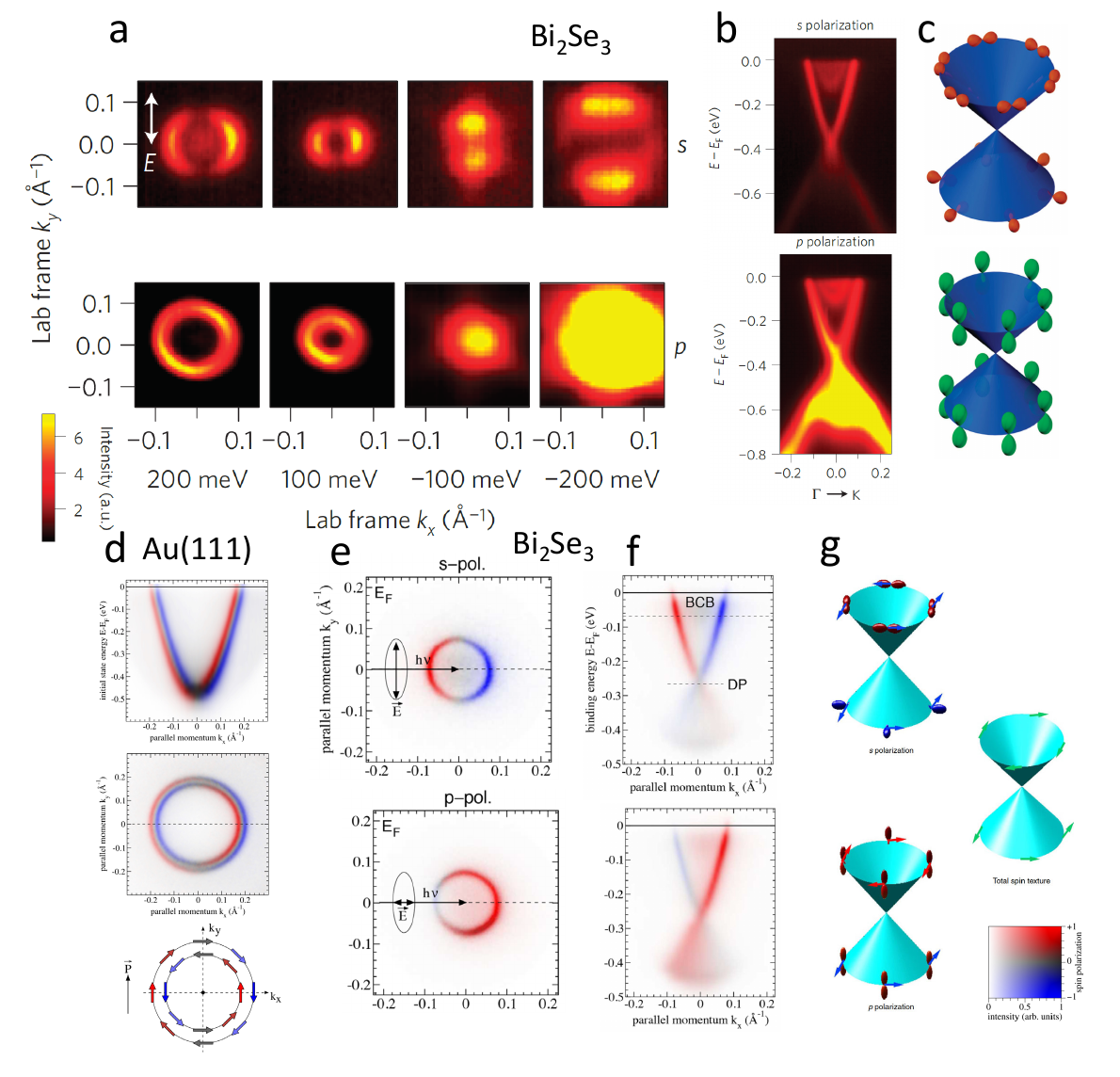}
	\caption{\label{FigBi2Se3spin}  
Orbital-dependent spin texture of Bi$_2$Se$_3$. The photoemission signal of (a) constant energy surfaces 
and (b) band dispersion of (c) the Dirac cone depends on the direction of the polarization vector ({\bf E} marked in panel (a)) \cite{CaoNP13}. With s-polarized light the orientation of in-plane $p_{x,y}$ orbitals can be determined using experimental geometries where a crystal mirror plane is along or perpendicular to the orbital axis. With p-polarized light 
predominantly $p_z$ orbitals are probed. (c) The result of the experiments in (a,b) is given in a sketch. 
(b) The intensity asymmetry for p-polarized light is due to the finite incidence angle of the light. 
(d-f) Modern spin-resolved photoemission instruments provide complete spin-resolved constant energy surfaces and band dispersions. This is shown for (d) the Rashba-split surface state on Au(111) which shows spin-momentum locking but is not topological.  (g) Spin-resolved ARPES has shown that the different p-orbital components have spin textures of opposite in-plane orientations  \cite{XieZspinNC14} which is demonstrated with state-of the art data in panels (e,f) \cite{MeyerheimTusche18}. The overall spin texture corresponds to the one of $p_z$ orbitals probed by p-polarized light. Panels a-c from \cite{CaoNP13}, d from \cite{Tusche15}, e, f from \cite{MeyerheimTusche18} and g from \cite{XieZspinNC14}.  }
\end{figure*}

As shown in  Fig.~\ref{FigBi2Se3} for Bi$_2$Se$_3$, these gapless TSSs are directly observed in ARPES as single band features linearly dispersing between the bulk valence and conduction bands across a topologically nontrivial  band gap   
of $\approx$\,0.3~eV, a value  {{\braun} exceeding} the energy scale of room temperature. 
The 2D nature of TSSs is verified by their lack of dispersion {{\braun}  with the electron wave vector $k_z$ perpendicular to the surface, which in ARPES is changed by the photon energy,}  in contrast to bulk bands which are permitted to disperse perpendicular to the surface. The  bulk band gap  is caused by strong spin-orbit coupling driving a band inversion transition at the bulk ${\rm\Gamma}$ point. The resulting single Dirac cone TSS exhibits a pronounced linear energy-momentum dispersion, $E(k_x,k_y)$\,$\propto$\,$|k|$, and a  chiral spin texture in which electron spins are locked perpendicular to their linear momenta, parallel to the surface plane. 
{{\braun} This spin texture was  first  measured by spin-resolved ARPES on Bi$_2$Te$_3$ \cite{HsiehspinNature09}.}
Extensive spin-resolved ARPES measurements 
\cite{PanVallaPRL11,SoumaPRL11,XuSYScience11,JozwiakPRL11}
identified and confirmed the predicted ground-state spin structure of the TSS in Bi$_2$Se$_3$, as shown in Fig.~\ref{FigBi2Se3spin}. These measurements revealed an in-plane spin orientation tangential to the Dirac cone with a left-handed chirality that reverses direction below the spin-degenerate  Dirac point. For Sb$_2$Te$_3$, the spin texture of the lower Dirac cone has been probed \cite{Pau12}. 

{{\braun} Bi$_2$Se$_3$ and Bi$_2$Te$_3$  are instrinsically n-doped so that the chemical potential is in the bulk conduction band. }
With ternary and quaternary compounds
  (Bi$_{1-x}$Sb$_x$)$_2$(Se$_y$Te$_{1-y}$)$_3$  
  a fully bulk-insulating or intrinsic 3D-TI phase is achieved  leading to surface-dominated conduction in electrical transport at room temperature. Moreover, the quaternary compound allows to control the energy of the Dirac point in the bulk band gap. ARPES is very useful to judge the effect of the doping, though surface band bending has to be taken into account  \cite{AnalytisPRB10}. 
   
\begin{figure*}[t]
	\centering
	\includegraphics[width=0.87\textwidth]{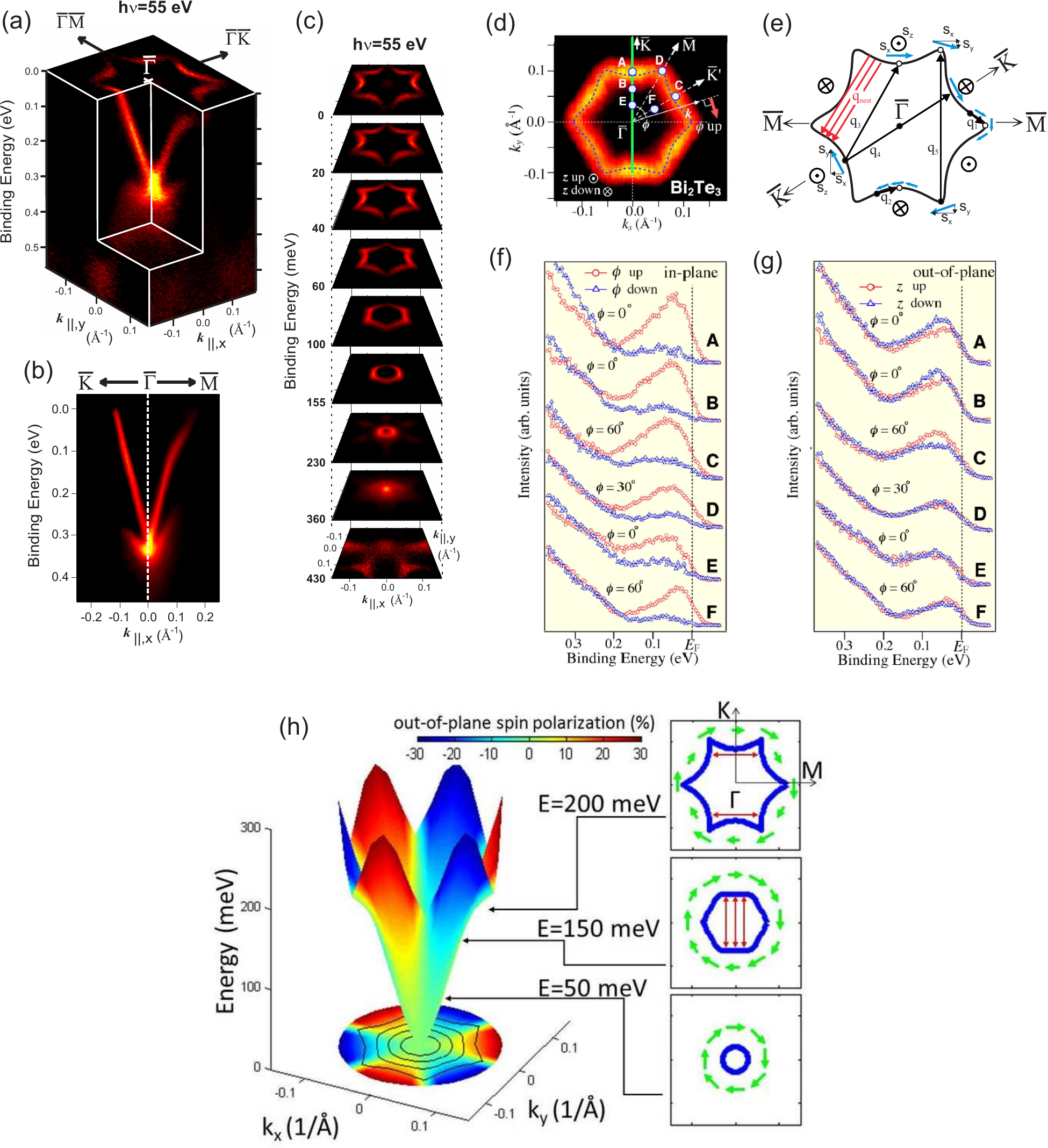}
	\caption{\label{FigWarping}  
Hexagonal warping in Bi$_2$Te$_3$. The warping was observed in ARPES at first by Chen et al. \cite{ChenYLScience09}.   (a-c) Band dispersions and constant-energy surfaces showing the evolution of a circular Dirac cone to a strongly warped one at the Fermi energy. The warping is outwards along the \dirGM\ direction. (d-g) Influence of warping on the spin texture. (d) Constant-energy surface depicting the positions in momentum space of (f,g) spin-resolved measurements for (f) in-plane and (g) out-of-plane spin. 
(e) Scattering channels based on the spin texture which affect the lifetime broadening. 
(h) Emergence of the out-of-plane spin polarization. 
Panels (a-c,e) from \cite{SanchezPRB14}, (d,f,g) from \cite{SoumaPRL11}, (h) from \cite{HasanPhysics09}. }
\end{figure*}

\begin{figure*}[t]
	\centering
	\includegraphics[width=0.9\textwidth]{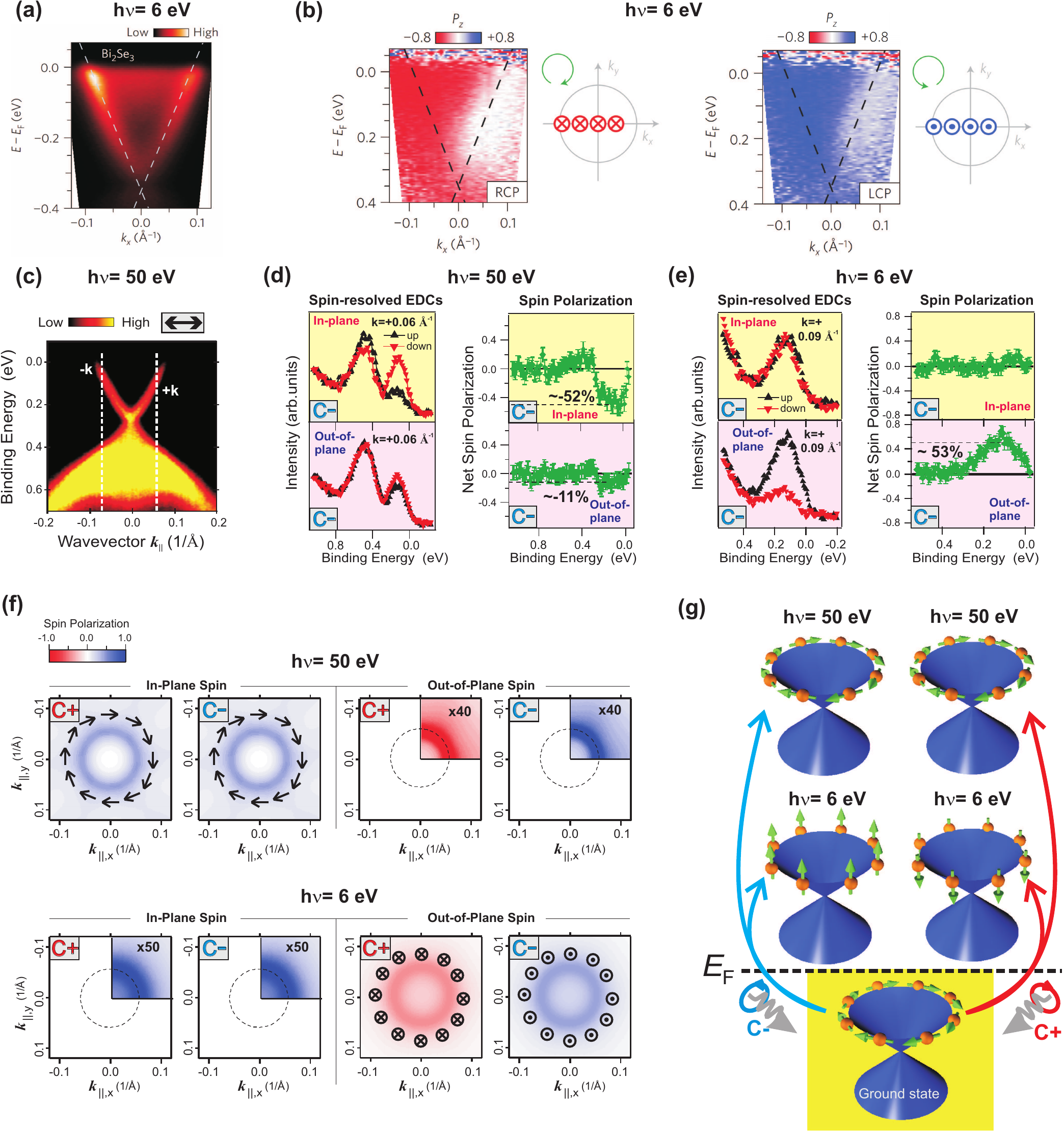}
	\caption{\label{FigBi2Se3spinmanip}  
Spin texture of photoelectrons from the Dirac cone of Bi$_2$Se$_3$. (a) Dispersion using 6 eV laser light and (b,c) out-of-plane spin polarization for opposite circular light polarization. (c) Dispersion at 50 eV photon energy for linear light polarization and corrsponding spin-resolved ARPES measurements for circular light polarization at (d) 50 eV and (e) 6 eV. (f) One-step-model photoemission calculations. At 50 eV the photoelectron spin polarization corresponds to the in-plane ground-state spin texture while at 6 eV the out-of-plane spin polarization reverses with the circular light polarization. (g) Simplified sketch of the experimental and theoretical results. Panels (a,b) from \cite{JozwiakNatPhys13}, panels (c-f) from \cite{SanchezPRX14}.  }
\end{figure*}

\subsection{{\normalfont\normalsize\bfseries II. B. Hexagonal warping}}

In a topologically non-trivial analogue of a free electron gas, a TSS would be entirely circular. In real systems this is often not the case, with the shape of the surface state able to reflect the underlying crystal symmetries~\cite{fu_hexagonal_2009}. The degree of warping is correlated to the localisation of the orbitals composing the band inversion and the topological state. In the Bi$_2$Se$_3$ material class, this manifests in a hexagonal warping of the constant energy contours of the surface state, usually developing apexes along the $\overline{\Gamma}$
-$\overline{K}$ direction. This warping, which is captured by the third-order correction to the Rashba Hamiltonian, opens an otherwise symmetry-forbidden out-of-plane spin {\braun component} along the  $\overline{\Gamma}$-$\overline{K}$ direction. 

Indeed, ARPES has discovered a snowflake-like, hexagonally-warped constant energy surface for  Bi$_2$Te$_3$ \cite{ChenYLScience09} (see Fig.~\ref{FigWarping}). This opens up possibilities for realizing a number of interesting phenomena, such as a surface quantum Hall effect under applied magnetic fields parallel to the surface, anisotropic surface scattering, or the formation of spin-density waves. 
As shown in Fig.~\ref{FigWarping} the hexagonal warping causes a strong anisotropy between the energy-momentum band dispersions of the TSS along the $\overline{\Gamma}$-$\overline{\mbox{K}}$ and $\overline{\Gamma}$-$\overline{\mbox{M}}$ directions of the surface Brillouin zone, as observed experimentally by ARPES. 

Following the first theoretical description of this effect, the influence of hexagonal warping on the band dispersion and spin texture of the TSS in Bi$_2$Te$_3$ was confirmed by spin-resolved ARPES measurements 
\cite{SoumaPRL11}. The experiments revealed a left-handed 3D spin texture with an out-of-plane spin component oscillating around the Fermi surface and an in-plane spin component tangentially following the snowflake contour \cite{ChenYLScience09} resulting from strong warping. The measured out-of-plane spin polarization was found to increase in magnitude with higher energy, to reverse its sign with an angular periodicity of $\pi$/3 around the hexagonally distorted constant-energy contours, and vanish along the $\overline{\Gamma}$-$\overline{\mbox{M}}$ direction due to mirror symmetry \cite{SoumaPRL11}. The influence of warping on the spin orientation was predicted   to alter the channels for quasiparticle scattering \cite{fu_hexagonal_2009}, and ARPES measurements \cite{SanchezPRB14} showed that it introduces an anisotropy in the lifetime broadening of the TSS, which is larger along the $\overline{\Gamma}$-$\overline{\mbox{K}}$ direction. 

Warping effects can also be induced in systems with significant $k_z$ dispersions, regardless of the strength of the coupling to underlying bulk symmetries. In these cases,  the surface electronic structure adopts unconventional band dispersions to avoid the energy cost associated with becoming resonant with the bulk bands (see, for example, both topologically trivial and non-trivial surface states in transition metal dichalcogenide (TMD) systems in Section VIII. Case Study: Transition Metal Dichalcogenides).

\subsection{{\normalfont\normalsize\bfseries II. C. Spin and orbital texture and final-state effects}}

It has frequently been suggested that circular dichroism in the angular distribution (CDAD) of ARPES encodes direct information regarding the spin texture of TSSs \cite{WangGedik11}. Circularly polarized light couples, however, mainly to the orbital angular momentum. This is, being employed, for example, in x-ray magnetic circular dichroism (XMCD) and in soft-x-ray absorption experiments where the selection rules for total angular momentum are exploited to access both the spin and orbital magnetic moment \cite{vdLaanReview14}. 
{\braun While one could argue that CDAD may give information on the spin texture where it coincides with the orbital texture \cite{DilReview19}, it would be desirable to conclude} from CDAD on the orientation of the orbital moment relative to the spin \cite{ParkSRPRL12,JungWPRB11}. The CDAD of the TSS of Bi$_2$Te$_3$ varies, however, strongly with the photon energies and changes sign several times \cite{ScholzCD13}. Similar behavior which indicates that the CDAD is a final-state effect was observed for Rashba-type spin textures in topologically trivial materials \cite{Arrala13,CrepaldiPRB14}. 
One-step model photemission calculations including the final state reproduced the sign changes in the CDAD and indicated also a behavior with photon energy different from that of the in-plane spin polarization \cite{ScholzCD13}.

A combination of complementary techniques is often required for a full, unambiguous characterisation. While initially spin-resolved ARPES experiments provided contradictory values for the in-plane spin polarization of the TSS, for example, in Bi$_2$Se$_3$ there is agreement that the measured in-plane spin polarization matches the theoretical predicton of $\sim50$\%\,
reduced from 100 \%\ due to spin-orbit interaction \cite{YazyevPRL10}. By studying the response of the measured spin properties to incident light and its polarizations, it is possible to gain further experimental insight into the spin texture of TSSs. Several experimental and theoretical works have investigated the effect of varying photon energy and light polarization on the chiral spin texture of TSSs, which has revealed that it can be decomposed into individual orbital contributions from different atomic layers. Through an analysis of the initial states and their orbital character, it has been demonstrated
that the total spin texture of the TSS in Bi$_2$Se$_3$ is a linear combination of different spin textures associated with each individual orbital component, namely the two opposite spin textures linked to the out-of-plane $p_z$ and in-plane $p_{x,y}$ orbitals, see Fig.~\ref{FigBi2Se3spin}. This phenomenon can be attributed to strong spin-orbit coupling, and the resulting spin textures coupled with the orbtal-specific components were coined `coupled spin-orbital textures'~\cite{zhang_spinorbital_2013, cao_mapping_2013}.
Since the dominant orbital character in the wave function of the TSS is $p_z$, this leads to a partial reduction in the magnitude of total spin polarization ($<$100\%).

This has been compared for the more isotropic Dirac cone of Bi$_2$Te$_2$Se and the warped Dirac cone of Bi$_2$Te$_3$ \cite{BentmannPRBL21}. 
For the isotropic Dirac cone, the measured momentum distribution of the photoemission intensity and the spin was found to qualitatively reflect the orbital composition and orbital-projected in-plane spin polarization, respectively. This was also observed for the in-plane spin-polarization of Bi$_2$Te$_3$. However, the out-of-plane spin polarization of Bi$_2$Te$_3$ has been found to change with photon energy \cite{SeibelPRB16,BentmannPRBL21} which was qualitatively confirmed by one-step photoemission calculations \cite{BentmannPRBL21}. 

Furthermore, due to the fact that the wave function of the TSS extends deep into the bulk (about 2 nm) and the orbital weights change with distance from the surface, the total spin texture can also be broken down into contributions from different atomic layers \cite{ZhuZH13}. 

As a result, a layer-dependent spin-orbital texture is observed, with a significant out-of-plane $p_z$-orbital contribution within the first five atomic layers (Se-Bi-Se-Bi-Se, or a quintuple layer, about 1 nm), while only a relatively small in-plane $p_{x,y}$ contribution is present except for the fifth atomic layer. It has been argued that this layer-dependence is responsible for the photon energy dependence of the spin-polarisation via quantum interference \cite{ZhuZH13,ZhuZH14}. 

A theoretical study suggested that the spin texture of the photoelectrons may differ from that of the TSSs due to a strong spin-flip effect during photoemission \cite{ParkLouie12}. This effect was later experimentally confirmed by a spin-resolved ARPES experiment on Bi$_2$Se$_3$ using a 6-eV laser source \cite{JozwiakNatPhys13}, see Fig.~\ref{FigBi2Se3spinmanip}. This raised the question whether the use of linearly or circularly polarized light will always rotate the spins of the Dirac cone in the photoemission process along the propagation direction of the incident light, controlling their orientation through the direction of light polarization. According to another spin-resolved ARPES experiment, the measured spin texture of photoelectrons reflects the total spin texture of the TSSs in the initial state (which for Bi$_2$Se$_3$ is in plane), indicating that the predicted final state spin-flip effect is a weak perturbation \cite{SanchezPRX14}. The relative importance of this effect ultimately depends on the final state the electron.
 
As shown in Fig.~\ref{FigBi2Se3spinmanip} for Bi$_2$Se$_3$, these experimental findings were consistent with one-step model photoemission calculations demonstrating that when vacuum ultraviolet photons with linear or circular polarization are used under different incident geometries, the magnitude and direction of the measured photoelectron spin polarization remain unchanged and  independent of the light polarization in the photon energy range 50 to 70\,eV.  Accordingly, the observed spin texture of the photoelectrons is consistent with the chiral, momentum-locked, in-plane spin texture of the TSS in the initial state, and the magnitude of total spin polarization corresponds to the theoretically predicted value in the ground state ($\approx$50\%). 
For very low photon energies of 6 eV, however, circularly polarized photons flip the photoelectron spins perpendicular to the surface and reverse the resulting out-of-plane spin texture of the photoelectrons with the sense of circular polarization \cite{JozwiakNatPhys13}. The conditions for light-induced spin manipulation at low photon energies are determined by dipole selection rules, which allow transitions into $s$-like final states. However, this condition is not fulfilled for typical photon energies in the vacuum ultraviolet spectral range, where relevant transitions from the Dirac cone occur primarily into lower symmetry $d$-like final states \cite{SanchezPRX14}. Therefore the original prediction is fulfilled at low laser energies despite the need for much higher photon energies to approximate free-electron-like final states. The final-state wave functions in both cases are the most symmetric and connected to the highest symmetry in the atomic case (i.e., an $s$-like final state wave function with angular momentum $l$=0), which is necessary for light-induced spin manipulation. Thus, very low or very high photon energies in the x-ray spectral range are required for light-induced spin manipulation \cite{SanchezPRX14}.

\subsection{{\normalfont\normalsize\bfseries II. E. Thin limit}}

\begin{figure*}[t]
	\centering
	\includegraphics[width=0.95\textwidth]{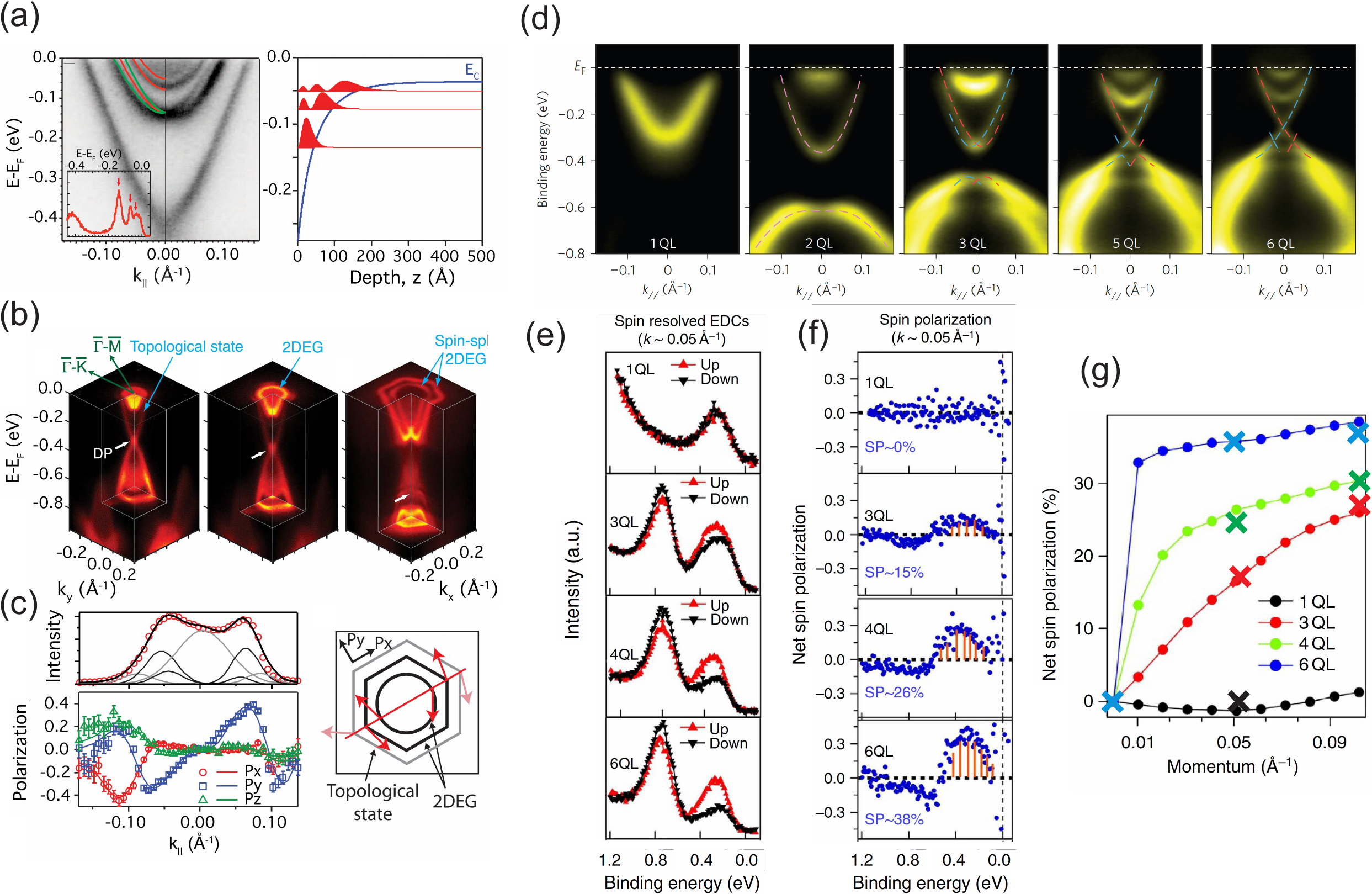}
	\caption{\label{Figthinlimit}  
Quantum-size effects in Bi$_2$Se$_3$. (a) Formation of a 2D electron gas (2DEG)  at the surface of a bulk crystal due to surface band bending. Quantum-well states are formed. (b) Formation of spin-split 2D Rashba states with increasing band bending. (c) Spin-resolved ARPES showing the Rashba splitting. 
(d-g) Topological phase transition as a function of thickness, in 5-atomic-layer units (quintuple layer, QL). (d) Dispersion and (e,f) spin polarization affected by interaction between top and bottom surfaces. (g) Summary of calculated (circles) and measured (crosses) spin polarization as a function of thickness. 
Panels (a-c) from \cite{KingPRL11}, (d) from \cite{ZhangYNP10},  (e-g) from \cite{NeupanetunnelingNC14}.   }
\end{figure*}

The effect of quantum confinement was initially investigated by ARPES experiments on Bi$_2$Se$_3$ bulk single-crystals, demonstrating that inducing a sufficiently strong band-bending is a feasible approach to achieve quantization of the bulk bands in the near surface region of a 3D TI \cite{KingPRL11}.
This is shown in   Fig.~\ref{Figthinlimit}(a-c) which also demonstrates that the quantized subbands possess a Rashba-type spin splitting due to the surface potential well. 

The weak van der Waals bonding of these compounds  leading to a relaxation of the requirement for lattice mismatching, allows for the growth of ultrathin films on various substrates by molecular-beam epitaxy (MBE), providing unprecedented control over thickness and composition. This advantage enables experimental investigations into the influence of chemical substitution, thickness, and dimensionality of topological properties. One example is the observation of a gap opening at the surface Dirac point in ultrathin TI films by ARPES, which demonstrates a topological phase transition driven by surface-to-surface tunneling between Dirac cones with opposite spin textures located on opposing film surfaces~\cite{ZhangYNP10}, see Fig.~\ref{Figthinlimit}(d). {\braun This has been studied by spin-resolved ARPES \cite{NeupanetunnelingNC14,LandoltPRL14}. As demonstrated} in Fig.~\ref{Figthinlimit}(e-g), this surface-to-surface coupling modulates the spin texture of the TSS with reducing thickness. 

\subsection{\normalfont\normalsize\bfseries III. Magnetic topological insulators} 

Since $\mathbb{Z}_2$ TIs are protected by time-reversal symmetry, a spontaneous magnetization can lead to a magnetic exchange gap  at the Dirac point. If the Fermi energy is aligned to the exchange gap, a quantum anomalous Hall effect can occur. Other desirable consequences include the topological magnetoelectric effect, axion insulator, dissipationless electronics and spintronics based on edge states, and the possibility for constructing topological quantum bits, see the review by Tokura et al. \cite{TokuraReview19}.

The ARPES perspective has focused on the verification and measurement of the exchange gap relative to the position of the Fermi energy as well as investigation of the hedgehog-type spin texture. The magnetic exchange gap was indirectly probed  by the observation of the quantum anomalous Hall effect  \cite{ChangCZ13}. From the thermal activation of dissipation-free transport, this gap is in the range of only 50-100\,$\mu$eV in V- and Cr-doped TIs, orders of magnitude lower than the principal limit set by the Curie temperature \cite{ChangCZPRL15}.

Spectroscopic investigations  relied on scanning tunneling spectroscopy  and ARPES. Early work suffered from a lack of evidence for the closing of the gap above the Curie temperature, a test that, due to stability issues, is easier in ARPES than in scanning tunneling spectroscopy. Another prerequisite for the exchange gap is a perpendicular magnetic anisotropy, see e.g. Ref. \cite{Henk12}, typically not present in \BiSe\-based systems.  
{\braun The reason for the latter is that the spin-orbit interaction has to be high enough so that the magnetocrystalline anisotropy overcomes the shape anisotropy, and this can be achieved by replacing Se by the heavier Te \cite{Rienks19}. The comparatively lower} spin-orbit interaction {\braun has also been suggested as reason why the \BiSe\ systems are} susceptible towards the opening of impurity-induced gaps at the Dirac point that are not of magnetic origin \cite{Rienks19}. 

Among early studies that fulfil some of these criteria, scanning tunneling spectroscopy of Cr$_{\rm 0.08}$(Bi$_{\rm 0.1}$Sb$_{\rm 0.9}$)$_{\rm 1.92}$Te$_3$ reported substantial disorder in the size of the exchange gap \cite{LeeI15} which might be responsible for the low operating temperature of the quantum anomalous Hall effect. ARPES of an insulating V-doped \BiSbTe\ epitaxial films showed that the Dirac point is degenerate with the bulk valence band and it was concluded that this limits the achievable temperature of the quantum anomalous Hall effect \cite{Li16}. No exchange gap could be found by ARPES down to 7~K \cite{Li16} and down to 1 K \cite{Golias21}. In both cases, the Dirac point was observed $\approx$50\,meV below the Fermi level~\cite{Li16,Golias21}. Cr-doped \BiSbTe\ was studied by ARPES at 12~K and no exchange gap was found also after increasing the chemical potential through surface $n$-doping by potassium~\cite{KimCK21}. 

\begin{figure*}[t]
	\centering
	\includegraphics[width=\textwidth]{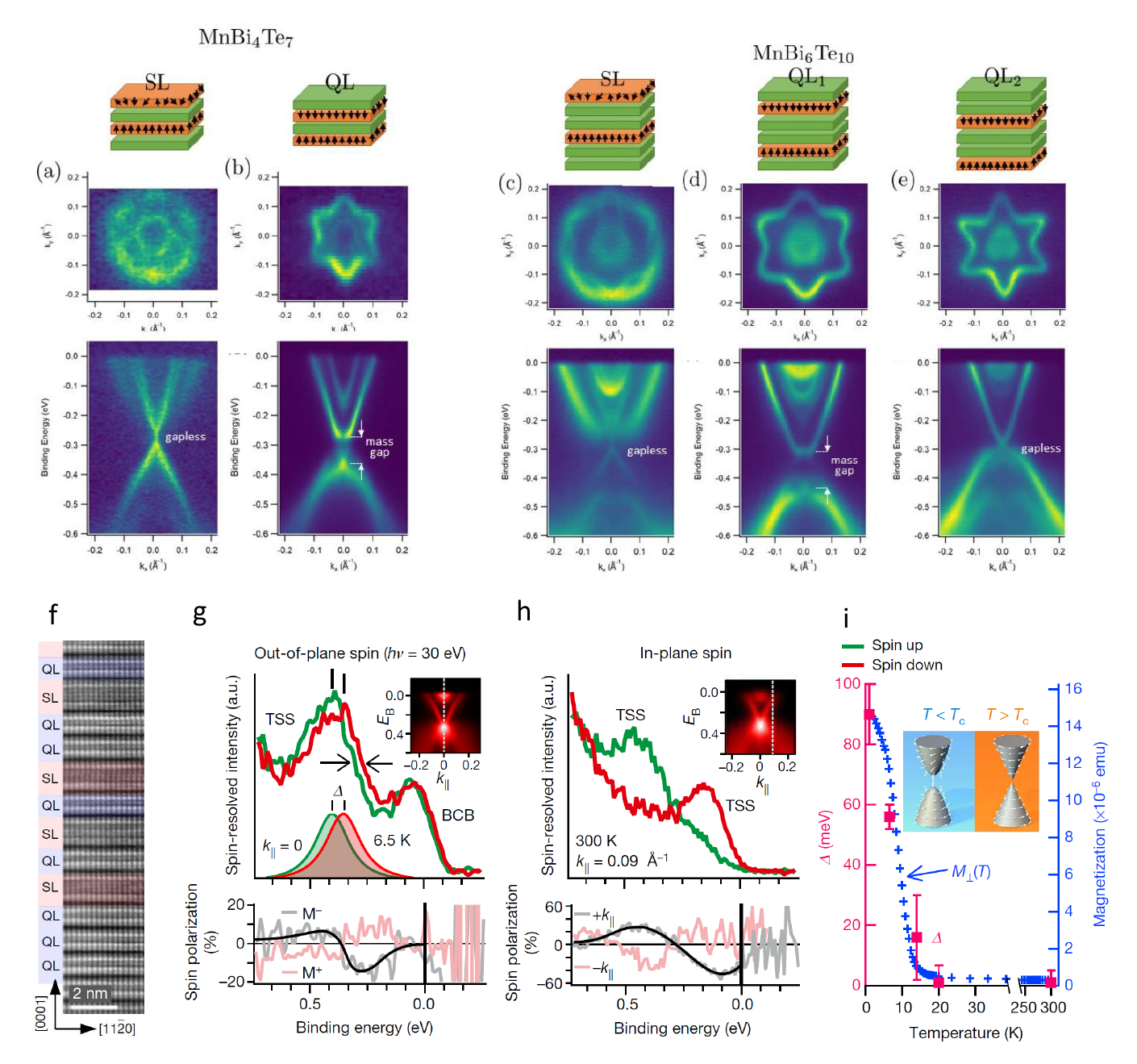}
	\caption{\label{FigMBT}  
Natural heterostructures MnBi$_2$Te$_4$/(\BiTe)$_{m}$. (a-e) Terminations with Bi$_2$Te$_3$ quintuple layers (QL) and MnBi$_2$Te$_4$ septuple layers (SL) have distinct features. (b,d) The Dirac cone appears gapped for 1 QL termination due to hybridization effects. (f) In thin films, the distribution of SL units is random, as seen in transmission electron microscopy. For 6\%\ Mn ($m=2$ to 3), the Dirac cone (topological surface state, TSS) shows {\braun (g) an exchange gap with out-of-plane spin texture which (h) away from the Dirac point develops into the spin-momentum locked in-plane spin texture.} (i) The gap closes at the Curie temperature {\braun and follows the macroscopic out-of-plane magnetization.} Panels a-e from \cite{Gordon19}, f-i from \cite{Rienks19}.}
\end{figure*}

Mn-doped \BiTe\ has perpendicular magnetic anisotropy \cite{HorYS10} and was later identified as consisting of a natural MnBi$_2$Te$_4$/(\BiTe)$_{m}$ heterostructure \cite{Rienks19}, similarly to its selenide counterpart \cite{Hagmann17,Hirahara17}. MnBi$_2$Te$_4$ contains a Mn layer in the center of a 7-layer unit (septuple layer).  The Mn concentration and the
number, $m$, of \BiTe\ quintuple layers can be related by x-ray diffraction \cite{Rienks19}. For 6\%\ Mn (i.e., $m$=2 to 3 \cite{Rienks19}) a gap at the Dirac point of 90 meV was measured by peak fitting in ARPES at 1K, and of 56 meV at 6.5 K by spin-resolved ARPES which closes at the Curie temperature~\cite{Rienks19}. This is shown in Fig.~\ref{FigMBT}. The closing of the gap at the Curie temperature has been confirmed in two other cases: For a gap of 70 meV at 8K for 3 and 5 net ferromagnetic quintuple layers of MnBi$_2$Te$_4$ \cite{Trang21} and for MnBi$_8$Te$_13$ (i..e, $m$=3) with septuple layer termination for a 28 meV gap at 7 K \cite{LuPRX21}. For MnBi$_6$Te$_{\rm 10}$ ($m$=2), three different terminations can be distinguished from the band dispersions \cite{Klimovskikh20}. Very similar results are obtained by Gordon \textit {et al.}~\cite{Gordon19} and Hu \textit {et al.}~\cite{HuDessau20} with mass gaps of the order of 100 meV when MnBi$_4$Te$_7$, MnBi$_6$Te$_{\rm 10}$, and MnBi$_8$Te$_{\rm 13}$ are terminated by one layer of \BiTe, {\braun see Fig.~\ref{FigMBT}.}

A single septuple layer of MnBi$_2$Te$_4$ has been characterized by ARPES as ferromagnetic trivial insulator \cite{Gong19} with a band gap $>$ 780 meV \cite{Trang21}. This septuple layer MnBi$_2$Te$_4$ becomes again topological when in a heterostructure of 4 quintuple layers of Bi$_2$Te$_3$ and another septuple layer MnBi$_2$Te$_4$ \cite{LiQ22}.

In stoichiometric bulk MnBi$_2$Te$_4$, the Mn layers couple antiferromagnetically to form an antiferromagnetic TI \cite{Otrokov19,Gong19}. The magnetic gap in   MnBi$_2$Te$_4$ has been the subject of controversy: {\braun Y. J. Chen et al. \cite{ChenYJPRX19} found 
a vanishing gap and exchange split bulk states.} It is possible that Mn-Bi intermixing \cite{Zeugner19} is responsible for differences in magnetic structure between different samples of MnBi$_2$Te$_4$ as well as those of MnBi$_2$Te$_4$/(Bi$_2$Te$_3$)$_m$. 

Epitaxial films of MnSb$_2$Te$_4$ are ferromagnetic TIs with Curie temperatures up to 50 K. The ferromagnetic order contrasts predictions of antiferromagnetism but is explained by Mn excess of the order of 5\%\. A slight $p$-doping prevents ARPES from measuring the gap at the Dirac point, but in scanning tunneling spectroscopy the gap is seen to vanish near the Curie temperature \cite{Wimmer21}. 
 
Among new materials, EuSn$_2$As$_2$ was characterized as antiferromagnetic TI by time-resolved ARPES above the Fermi level \cite{LiHPRX19}. The Weyl semimetal phase of MnSb$_2$Te$_4$ \cite{Murakami19} and MnBi$_2$Te$_4$ \cite{LiJ19}, has not yet been reached without external magnetic field. FeBi$_2$Te$_4$ has recently been synthesized \cite{Saxena20} and CrBi$_2$Te$_4$ and CrBi$_2$Te$_2$Se$_2$ have been predicted as magnetic TIs \cite{Petrov21}. 

\subsection{{\normalfont\normalsize\bfseries IV. Topological crystalline insulators and weak topological insulators}}

\begin{figure*}[t]
	\centering
	\includegraphics[width=0.76\textwidth]{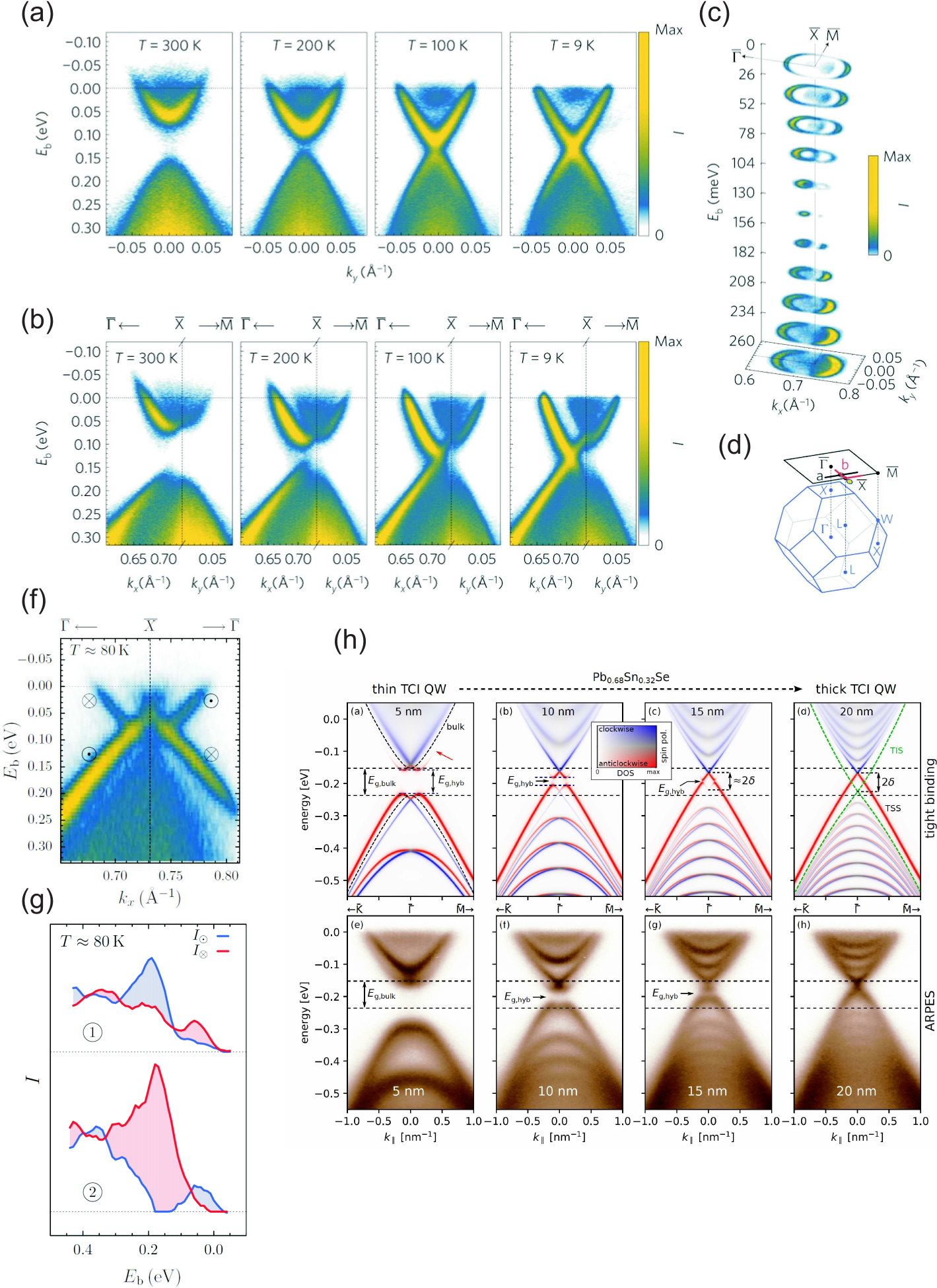}
	\caption{\label{FigTCIs}  
Topogical crystalline insulator (Pb,Sn)Se. (a,b) Temperature-driven topological phase transition from a trival (300 K) to a topolical crystalline insulator (low temperature). (c) Constant-energy surfaces showing the Dirac cone in the topological phase. (d) Corresponding surface and bulk Brillouin zone. (f) Dispersion and (g) Spin-resolved ARPES of the in-plane spin texture of the topological state showing spin-momentum locking. (h) Thickness-dependent topological phase transition in (Pb,Sn)Se. 
Panels (a-d) from \cite{DziawaNM12},  (f,g) from \cite{WojekPRB13},  (h) from \cite{RechcinskiAFM21}.
 }
\end{figure*}

In contrast to the Dirac cones in strong TIs which are protected by time-reversal symmetry, a second class, known as topological crystalline insulators (TCIs) exhibits Dirac cones protected by the crystal symmetry itself, see the review by Ando and Fu \cite{AndoFu15}.  An example is the rock-salt structured Pb$_{1-x}$Sn$_x$(Se,Te)  series, see Fig.~\ref{FigTCIs}.  Dirac-cone TSSs have been observed by ARPES for SnTe 
\cite{TanakaNP12}, Pb$_{0.6}$Sn$_{0.4}$Te \cite{XuSYNC12}, and the Pb$_{1-x}$Sn$_x$Se series \cite{DziawaNM12}. They possess a chiral spin texture and spin-momentum locking. Since TCIs are protected by individual crystal symmetries and bear a weak topological index, the quantum phase transition in TCIs {\braunb is vulnerable} to external perturbations
\cite{r49B1,r47B1}.

For TCIs, the decisive role of the crystal symmetry renders the topological protection dependent on the specific crystal face \cite{r25B2}. 
The topological invariants allow for an even number of Dirac cones 
\cite{r13B2,r25B2}. 
{\braunb (Pb,Sn)Se and (Pb,Sn)Te
represent} mirror-symmetry protected TCI phase with fourfold valley degeneracy
\cite{r13B2,r26B2},
in which the trivial-to-TCI phase transition is reached for sufficiently large Sn contents and with reducing temperature \cite{TanakaNP12,XuSYNC12,AndoFu15,DziawaNM12}.
Upon cooling, the lattice contracts and the enhanced orbital overlap leads to an inverted bulk band gap which, via bulk-boundary correspondence, gives rise to an even number of spin-polarized Dirac cone surface states {\braunb with spin-momentum locking}, as confirmed experimentally by ARPES \cite{DziawaNM12} and spin-resolved ARPES \cite{WojekPRB13}, {\braunb see Fig.~\ref{FigTCIs}(a-f).} Since the TSSs in TCIs are only protected by {\braunb crystal} symmetry, they are an ideal platform for manipulating the topological properties. Due to their high sensitivity to external perturbations, the topological phase transition can be controlled by varying pressure, hybridization in ultrathin film geometries, magnetic interactions, or by breaking of mirror symmetry by strain, electrostatic fields, and ferroelectric lattice distortions
\cite{r26B2,AndoFu15,r74B2,r75B2,r76B2,r77B2}.
These properties offer an advantage for topology control not available in   $\mathbb{Z}_2$-TIs, such as to open a gap at particular Dirac points by quantum confinement or unidirectional strain. For SnTe  a topological phase transition from a TCI to a $\mathbb{Z}_2$-TI by a lattice distortion has been predicted  \cite{r77B2}. 
Bi-doping of Pb$_{1-x}$Sn${_x}$Se~(111) epilayers induces a quantum phase transition from a TCI to a strong $\mathbb{Z}_2$-TI by lifting the fourfold valley degeneracy via a sublattice shift and rhombohedral distortion along the [111] direction. The phase transition was confirmed experimentally by ARPES through the number of gapless Dirac cones \cite{MandalNC17}.

Up to now, most experimental work has been performed on $p$-type bulk single crystals exploiting the natural (001) cleavage plane of the IV-VI compounds, whereas for other surface orientations and actual devices, epitaxial TCI film structures are required. To this end, the (111) orientation is particularly interesting due to the polar nature of its surface. For instance, full control of the Fermi level and carrier concentration {\braunb leading to a giant Rashba effect} was achieved for Bi- and Sb-doped TCI Pb$_{1-x}$Sn${_x}$Te~(111) films  \cite{Volobuev17}.  Similarly, control of both the topological phase transition and the structure inversion asymmetry additionally   was achieved for TCI quantum wells in ultrathin Pb$_{1-x}$Sn${_x}$Se films grown on Pb$_{1-y}$Eu$_{y}$Se {\braunb barriers \cite{RechcinskiAFM21}, see Fig.~\ref{FigTCIs}(g).}

{\braunb The TCIs presented above depend on a symmorphic mirror plane. Symmorphic symmetries preserve the
origin whereas nonsymmorphic spatial symmetries translate the origin by a rational fraction of the lattice vector.  This leads to new classes of topological crystalline insulators \cite{LiuCXPRB14}, for example, such that lead to a symmetry-protected surface Dirac point that is not pinned to a specific point \cite{FangFuPRB15}. 
In time-reversal invariant nonsymmorphic TIs, instead of surface Dirac fermions, the appearance of surface fermions with hourglass-shaped dispersion was predicted where two pairs of branches switch their partners \cite{WangZNature16}. This behavior was predicted for  KHgX (X = As, Sb, and Bi) \cite{WangZNature16}
and was demonstrated by ARPES for the (010) surface of KHgSb \cite{MaSA17,LiangPRB17}.}

Besides TCIs there is another topological class which leads to TSSs only on particular suefaces. These are weak TIs which are not protected by crystal symmetries but by time-reversal symmetry in the same way as strong topological insulators. They can be consdered as 2D quantum spin Hall insulators stacked in 3D. The side surfaces are often difficult to prepare in expriment which renders their confirmation difficult despite several theoretical predictions, e.g. for PbTe/SnTe superlattices \cite{YangG14}. Bi$_1$Te$_1$, which is a superlattice of 2 quintuple layers of Bi$_2$Te$_3$ and one Bi bilayer, has been predicted to be a dual TI, i.e., in this case a TCI and a weak TI. The TSS on the top surface, protected by mirror symmetry, has been confirmed by ARPES but not the TSS on the side surface protected by time-reversal symmetry following the picture of the stacked 2D quantum Hall insulator \cite{Eschbach17}. Another prediction for a weak TI concerns $\beta$-Bi$_4$I$_4$ \cite{LiuBiI16}. $\beta$-Bi$_4$I$_4$ consists of stacked molecular chains and shows in ARPES a 1D Dirac cone surface state \cite{Autes16}. The system has  naturally cleavable top and side surfaces. By spatially resolved ARPES, a quasi-1D Dirac cone was found at the side surface (100) whereas the top surface (001) shows no TSS \cite{Noguchi19}.

\begin{figure*}[t]
	\centering
	\includegraphics[width=\textwidth]{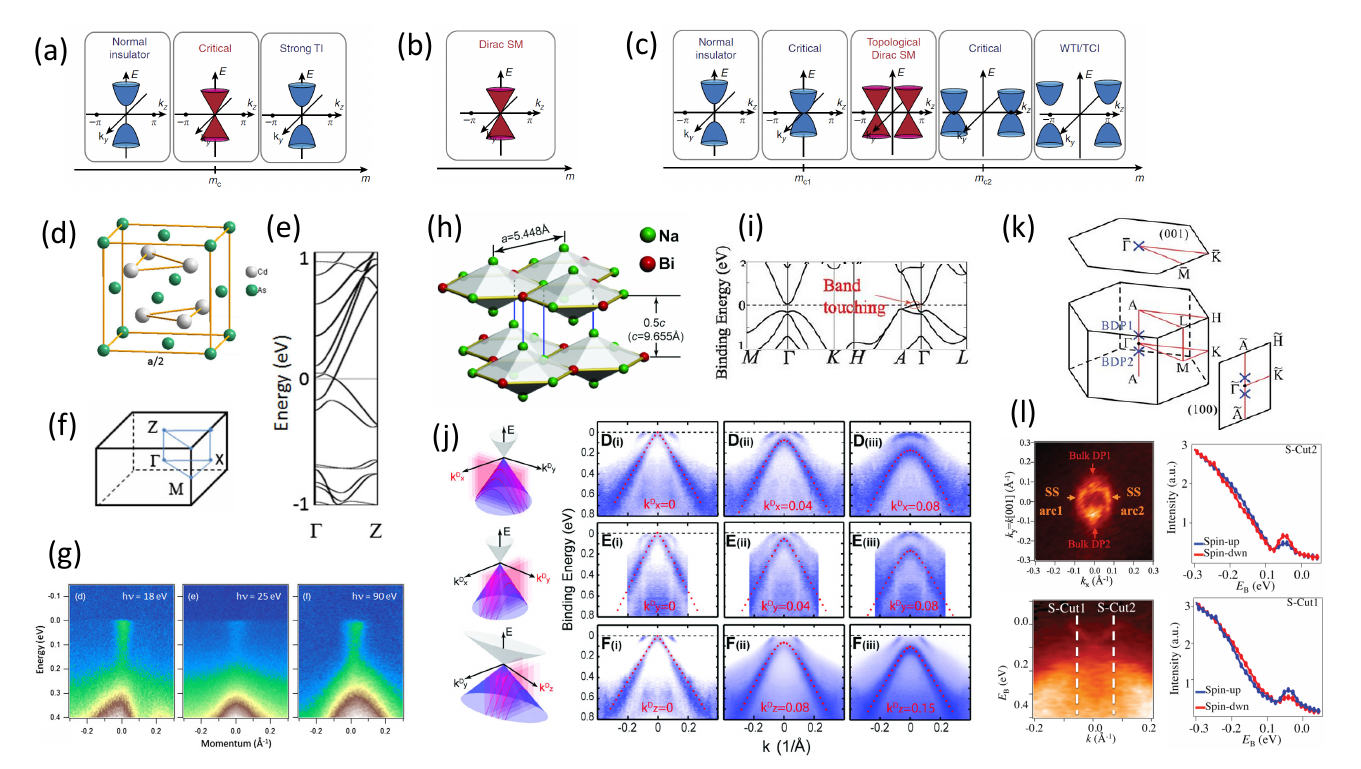}
	\caption{\label{FigDMS}  
3D Dirac semimetals Cd$_3$As$_2$ and Na$_3$Bi. Phase transition upon variation of control parameter $m$ leading to  a 3D Dirac semimetal by (a) accidental band crossing or with  (b) additional protection by rotational symmetry. (c) Topological 3D Dirac semimetal by band inversion and protection by rotational symmetry.  
(d-g) Cd$_3$As$_2$ of $C_3$ symmetry shows the (e) predicted 3D Dirac points in (g) ARPES along the $\Gamma$-$Z$ direction. 
(h-l) Na$_3$Bi of $C_4$ symmetry shows (i) 3D Dirac points along the $\Gamma$-$A$ direction which are seen in (j) ARPES by variation of the momentum along all 3 directions. 
(k) The projected Dirac points are separated at the (100) surface where (l) two Fermi arcs are observed which exhibit opposite spin polarization. Panels (a-c) from \cite{yang_classifcation_2014}, panels (d-g) from \cite{BorisenkoPRL14}, panels (h-j) from  \cite{LiuZKScience14}, and panels (k,l) from \cite{XuSYScience15}. 
 }
\end{figure*}

\subsection{{\normalfont\normalsize\bfseries V. Dirac semimetals}}

3D Dirac semimetals emerge at the quantum critical point of a topological phase transition between a trivial insulator and a $\mathbb{Z}_2$-TI \cite{WangZPRB12}. Closing the bulk band gap in a TI gives rise to a 3D Dirac semimetal phase that can be viewed as the bulk analog of graphene, which is a 2D Dirac semimetal. {\braunb For  Bi$_2$Se$_3$, for example,}  , moving across the topological quantum phase diagram would be done theoretically by continuous control of the spin-orbit coupling strength \cite{ref43,ref66}. 

A linear bulk band crossing exactly at the quantum critical point between topologically trivial and non-trivial phases could be accurately described as a Dirac semimetal, with a single fourfold degenerate 3D Dirac point and linear energy dispersions along all three orthogonal momentum directions. Appropriate tuning of the chemical potential in this case would result in a point-like Fermi surface. However, this crossing point could be destroyed by minute perturbations, pushing the Dirac semimetal {\braunb to} either side of the topological transition to produce a finite gap, see Fig.~\ref{FigDMS}(a). Experimentally realized topological Dirac semimetals instead employ intrinsic rotational crystalline symmetries to prevent hybridisation between crossing bands,  ensuring the formation of robust bulk Dirac points.

Yang and Nagaosa \cite{yang_classifcation_2014} presented a general framework for classifying stable 3D Dirac semimetals in systems characterized by the presence of time-reversal, inversion, and uniaxial rotational symmetries. Two distinct classes of 3D Dirac semimetals are identified in their study. In one class, the Dirac semimetal exhibits a single Dirac point at a time-reversal invariant momentum on the rotation axis where the band crossing is ensured by the lattice symmetry, see Fig.~\ref{FigDMS}(b). In contrast, the other class of Dirac semimetals  features a pair of Dirac points on the rotation symmetric about the time-reversal invariant momentum, see Fig.~\ref{FigDMS}(c). Here, the Dirac points are formed through band inversion and this class is accompanied by a quantized topological invariant and commonly referred to as 3D topological Dirac semimetals. A further spectroscopic signature is the presence of Fermi arc states connected between bulk Dirac points in topological DSMs \cite{WangZPRB12,WangZPRB13} which were at first discussed  in the context of 3D Weyl semimetals \cite{Wan-PRB-2011}. When the Dirac point acquires a mass gap because of symmetry breaking, the 3D topological Dirac semimetal can turn into either a 3D strong topological insulator or a 3D TCI.

Experimentally, Cd$_3$As$_2$ \cite{BorisenkoPRL14,AliInorgChem14,NeupaneCdAsNC14} and Na$_3$Bi \cite{LiuZKScience14,XuSYDSMarXiv13} were the first 3D topological Dirac semimetals that were confirmed by ARPES. A highly linear bulk band crossing forming a 3D Dirac cone projected at the Brillouin zone center of the (001) surface of Cd$_3$As$_2$ have been observed down to photon energies of about 22 eV \cite{NeupaneCdAsNC14}. 
A systematic confirmation along all three orthogonal momentum directions followed for   Cd$_3$As$_2$ and Na$_3$Bi  \cite{BorisenkoPRL14,AliInorgChem14,NeupaneCdAsNC14,LiuZKScience14,XuSYScience15}, see Fig.~\ref{FigDMS}. 
ARPES and spin-resolved ARPES experiments on Na$_3$Bi  revealed a double Fermi arc surface structure which is spin polarized and  confirms the topological protection \cite{XuSYScience15}, see Fig.~\ref{FigDMS}(l). 
 It has been proposed that Cd$_3$As$_2$ and Na$_3$Bi would transition into a TI state if they are distorted \cite{WangZPRB12,WangZPRB13}. This is another interesting aspect which concerns the protection of the 3D Dirac points by the symmetry of the lattice, potentially enabling the switching on and off of their contribution to transport by external perturbations, such as strain or  other structural distortions \cite{WangZPRB12,WangZPRB13}. 
 Such a topological phase transition would ideally require an intrinsic mechanism by which to induce a structural distortion to be easily accessible by ARPES. Au$_2$Pb is thought to be a candidate for this purpose, as   it naturally undergoes several structural phase transitions with reducing temperature \cite{SchoopPRB15}. 

Breaking inversion or time-reversal symmetry, or both, in a 3D Dirac semimetal is predicted to be a fruitful platform for the {\braunb observation of} topological phase transitions. A direct consequence is that the 3D Dirac points will split into Weyl points of opposite chirality in momentum space.

\subsection{{\normalfont\normalsize\bfseries VI. Nonmagnetic Weyl semimetals}}

3D Weyl semimetals are a nontrivial state of matter with peculiar low-energy excitations and   transport properties \cite{Wan-PRB-2011, Balents-Viewpoint-2011,Balents-Weyl-Multilayer-2011,Hosur-transport-Weyl-2013, Bernevig-Weyl-2015,Armitage2018,Narang-NatPhys-2021}. They combine a quasi-relativistic dispersion which is linear in momentum in all three spatial directions with topological protection \cite{Soluyanov-Nature-2015,Hasan-NatComm-Weyl-2015,Bansil-RMP-2016}, and exhibit extremely high carrier mobilities which could enhance the efficiency of future electronic devices \cite{Yang-PRB-Weyl-Applications-2011,Sun2016}. Due to their nontrivial topology, 3D Weyl semimetals host spin-polarized topological Fermi arcs at the surface while being metallic in the bulk \cite{Armitage2018,Narang-NatPhys-2021,Soluyanov-Nature-2015,Hasan-NatComm-Weyl-2015}. They offer a rich platform for the realization of an unusually large magnetoresistance \cite{Felser-NatPhys-Transport-Weyl-2015} and  applications in future electronics and spintronics \cite{Sun2016}. 

A 3D Weyl semimetal phase arises from breaking time-reversal symmetry or inversion symmetry in a 3D Dirac semimetal. 
If in addition the spin-orbit interaction is sufficiently strong such that the bulk bands remain in the inverted regime and gapless, each fourfold degenerate bulk Dirac point in a topological 3D Dirac semimetal will split into two 3D Weyl points of opposite chirality. Hence,
for each original Dirac point in the 3D Dirac semimetal, breaking any of the two symmetries leads to a pair of twofold degenerate 3D Weyl cones that are topologically protected and exhibit quasi-relativistic energy dispersion in 3D momentum space. 
Consequently, the spin-polariszed Fermi arcs connecting bulk Dirac points no longer form closed loops in the 3D Weyl Semimetal phase. Instead, the surfce Fermi arcs connect pairs of 3D Weyl points with opposite chirality, and form open loops when projected on the surface Brillouin zone. The Weyl nodes themselves act as channels through the bulk of the crystal to the opposite surface, thus ensuring a globally closed contour of Fermi arcs. In line with this, Weyl points arise as pairs with opposite chirality. Indeed, the chirality associated with each Weyl point is a topological property defined as a topological charge given by a nonvanishing Chern number, i.e., topological invariant, which means that the Weyl cones can only be moved in energy and momentum space, or tilted, rather than being gapped by external perturbations.

\begin{figure*}[t]
	\centering
	\includegraphics[width=\textwidth]{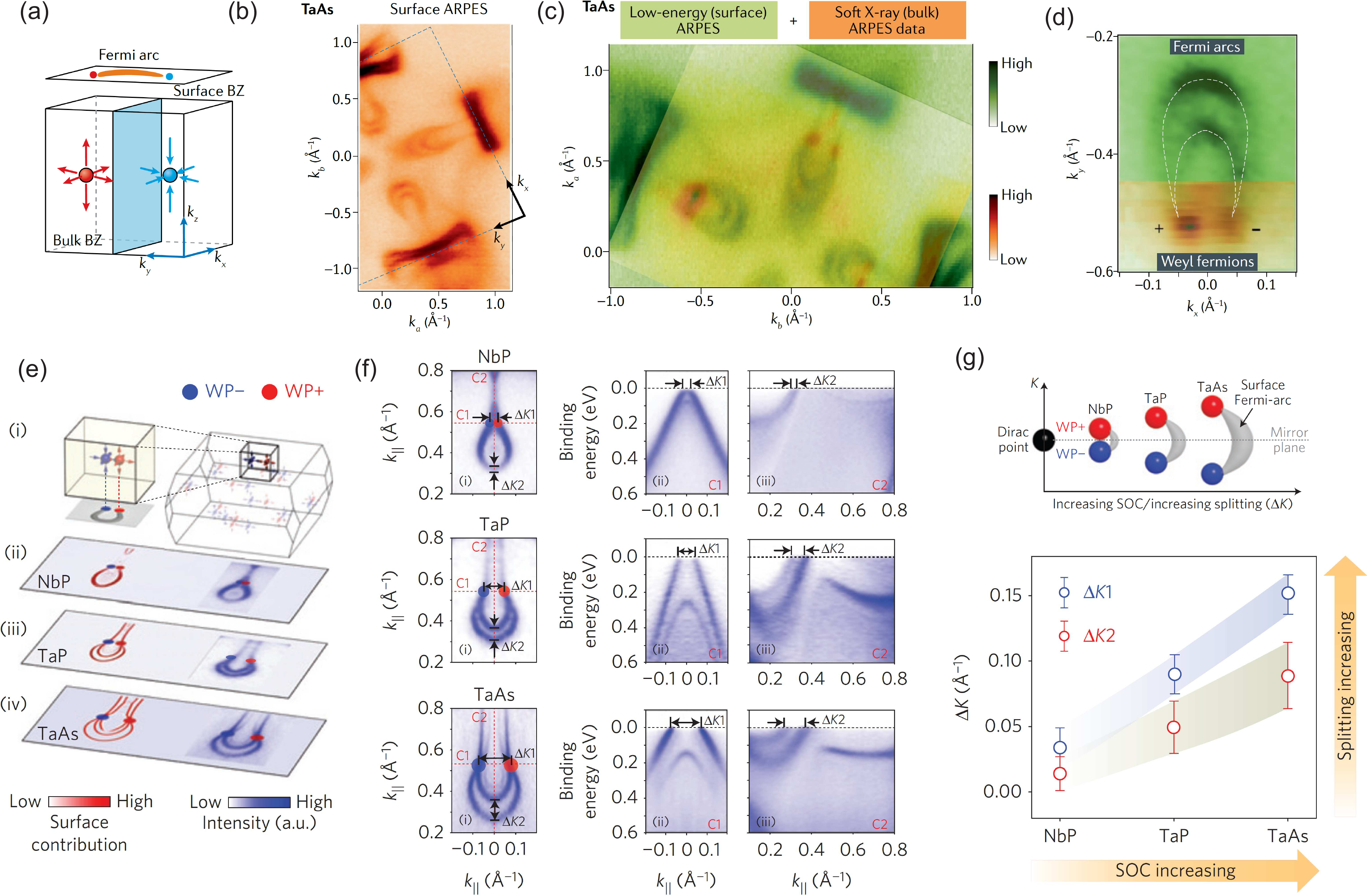}
	\caption{\label{FigWeylSM}  
3D Weyl semimetals. (a) Schematic bulk and surface Brillouin zone (BZ)  depicting the projection of 3D Weyl points of opposite chirality on the surface along with the surface Fermi arc. (b-d) Fermi surface by ARPES of TaAs measured (b) with surface-sensitive ARPES. (c,d) The corresponding superposition with bulk-sensitive soft-x-ray ARPES (red in c and d) depicting the bulk Weyl points (WP) projected onto the surface along with the surface Fermi arcs (green in c and d). (d) Zoom in on the region of the surface Fermi arcs where the positive and negative signs indicate the  chiralities of the Weyl points. (e-g) Comparison of different pnictide Weyl semimetals. From NbP through TaAs the splitting between the Weyl points and the length of the surface Fermi arcs increases. This splitting increases with the spin-orbit coupling. Panels (a-d) from \cite{XuTaAsScience15} and (e-g) from \cite{LiuZKNM16}.
  }
\end{figure*}

In the case of 3D Weyl semimetals, a minimal setting corresponds to the situation of a 3D Dirac semimetal, but in which time-reversal symmetry is broken while inversion symmetry preserved. When only inversion symmetry is present, a 3D Weyl point at $k$ must necessarily be accompanied by a partner node at -$k$ that carries the opposite topological charge at the same energy. This configuration allows for a minimum number of Weyl points with opposite chirality. In contrast, when only inversion symmetry is broken, time-reversal symmetry requires that these two Weyl points at opposite momenta carry the same topological charge. In this situation, because the total topological charge within the entire Brillouin zone should vanish, another pair of Weyl points with the opposite topological charge   is additionally required to achieve charge neutrality. A natural consequence is that the Weyl nodes in 3D Weyl semimetals without inversion symmetry must   necessarily exist in multiples of four. This fact also provides a convenient way to investigate a variety of interesting effects associated with topological charge redistribution, such as the efficient generation of optically-driven spin-polarized currents from highly tilted Weyl cones or other photoinduced responses like the photovoltaic chiral magnetic effect or the gyrotropic magnetic effect. 

The first materials confirmed experimentally were inversion-symmetry breaking 3D Weyl semimetals of the TaAs family, primarily by ARPES \cite{YangLXTaAsNP15,XuTaAsScience15,LiuZKNM16}. Predicted materials with a stable 3D Weyl semimetal phase of this family of compounds included TaAs, TaP, NbAs, and NbP. These materials belong to the class of topological semimetals known as type-I Weyl semimetals, which exhibit a point-like Fermi surface at the Weyl nodes.   As shown in Fig.~\ref{FigWeylSM}, surface Fermi arcs connecting pairs of 3D Weyl points projected onto the surface Brillouin zone and the linear energy-momentum dispersion of bulk states were systematically identified by extensive ARPES measurements  \cite{YangLXTaAsNP15,XuTaAsScience15,LiuZKNM16}. In addition to photon-energy dependent ARPES, the surface and bulk character of the different states were independently confirmed by low-energy surface-sensitive ARPES and soft-x-ray bulk-sensitive ARPES, the latter of which enabled the observation of the 3D Weyl nodes in isolation. 

One important aspect  facilitating the experimental verification of topological order is the influence of increasing spin-orbit coupling strength on the electronic properties of these compounds. As seen in the comparison between ARPES measurements of the Fermi surface and band dispersions of NbP, TaP, and TaAs shown in Fig.~\ref{FigWeylSM}, increasing the spin-orbit strength increases both the separation between the Weyl nodes in momentum space and the splitting of the band dispersions, which causes a splitting of the Fermi arcs while retaining the overall topology of the Fermi surface \cite{LiuZKNM16}. While this observation by itself provides additional evidence for the Weyl semimetal state, it also allows to experimentally identify groups of Weyl nodes with zero net chirality, establishing an independent criterion for the experimental verification of topological order. 

By combination of bulk-sensitive soft-x-ray ARPES, spin resolution and CDAD. it has been possible to observe large spin and orbital-angular-momentum polarizations in TaAs \cite{Uenzelmann}. From the winding of the orbital angular momentum, the Weyl points have been confirmed as Berry flux monopoles \cite{Uenzelmann}.


 Type-II Weyl semimetals   exhibit strongly tilted 3D Weyl cones in energy-momentum space, resulting in a Fermi surface where the upper and lower cones form electron and hole pockets that connect at the Weyl points. In consequence, type-II Weyl semimetals undergo perculiar Lifshitz transitions. 
 The original prediction included the layered material WTe$_2$, known for its large, nonsaturating magnetoresistance, as well as MoTe$_2$. Moroever, ARPES found also evidence for a type-II Weyl semimetal in TaIrTe$_4$ \cite{KoepernikPRB16,HauboldPRB17} and LaAlGe \cite{XuSYLaAlGe17}. 

\subsection{{\normalfont\normalsize\bfseries VII. Magnetic Weyl semimetals and kagome systems}}

The alternate route towards the creation of topological Weyl semimetals from Dirac semimetals is to instead break time-reversal symmetry by the application of an external magnetic field or through a spontaneous magnetization.  In the ferromagnetic Heusler compound  
Co$_2$MnGa, linear bulk band dispersions were observed, which intersect in a nodal line near the Fermi energy \cite{Belopolski19} and are spin polarized \cite{SumidaCM20}. Another such crossing at higher binding energy is spanned by a surface state in the shape of a drumhead supporting the assignment of the system to a Weyl semimetal \cite{Belopolski19}.

{\braun YbMnBi$_2$ is a magnetic system with a canted antiferromagnetic structure. 
By comparison to the control system EuMnBi$_2$, Weyl points and a Fermi surface arc have been observed and the system was classified as a TRS-breaking type-II Weyl semimetal \cite{BorisenkoNC19}. }

Insulating kagome materials can exhibit spin liquid phases as a result of magnetic frustration. The same interferences determine the electronic structure which is particularly interesting for metallic systems. With spin-orbit coupling and spontaneous magnetization, a two-dimensional kagome lattice leads to a 2D Chern insulator phase with quantized anomalous Hall conductance \cite{XuGPRL15}. The resulting 2D Dirac cones have been observed by ARPES in Fe$_3$Sn$_2$ (with Fe$_3$Sn kagome layers between Sn layers) and in FeSn \cite{KangMNM20}. 
{\braun  A gap in the 2D Dirac cone in Fe$_3$Sn$_2$ observed by ARPES has been assigned to spin-orbit interaction} \cite{YeLNature18}. Depending on the stacking of the kagome layers, the system can transform into a magnetic 3D Weyl semimetal \cite{LiuDF19}.

{\braunb The kagome magnet $R$Mn$_6$Sn$_6$ with rare-earth element $R$, in which Mn forms the kagome layer, is promising for the realization of different quantum phases due to a variety of magnetic structures.} In the system YMn$_6$Sn$_6$ a Dirac cone dispersion was seen by Li et al. \cite{LiM21}. 

Another signature {\braunb of the kagome lattice} is a flat (i.e. minimally dispersive) band. Such a flat band is observed $\approx$0.25 eV below the Fermi level in FeSn \cite{KangMNM20}, and similar flat bands appear in paramagnetic CoSn near the Fermi energy \cite{LiuZ20,KangMNC20}. These observations are interesting since a flat band on the kagome lattice carries a finite Chern number enabling a fractional quantum Hall state. 

\subsection{{\normalfont\normalsize\bfseries VIII. Case Study: Transition Metal Dichalcogenides}}

With time it is becoming apparent that non-trivial band topology, along with Dirac and Weyl nodes, are far from contained to the handful of compounds within which the first examples were found. 
Conversely, topological band properties are ubiquitous, generic features of band structures found across nature. Indeed, computational studies  tasked with scouring the electronic structures of all existing solids for non-trivial band topology predict that more than one quarter of all existing solids have an inverted band gap somewhere within their electronic structures~\cite{vergniory_complete_2019}.

\begin{figure*}[t]
	\centering
	\includegraphics[width=0.9\textwidth]{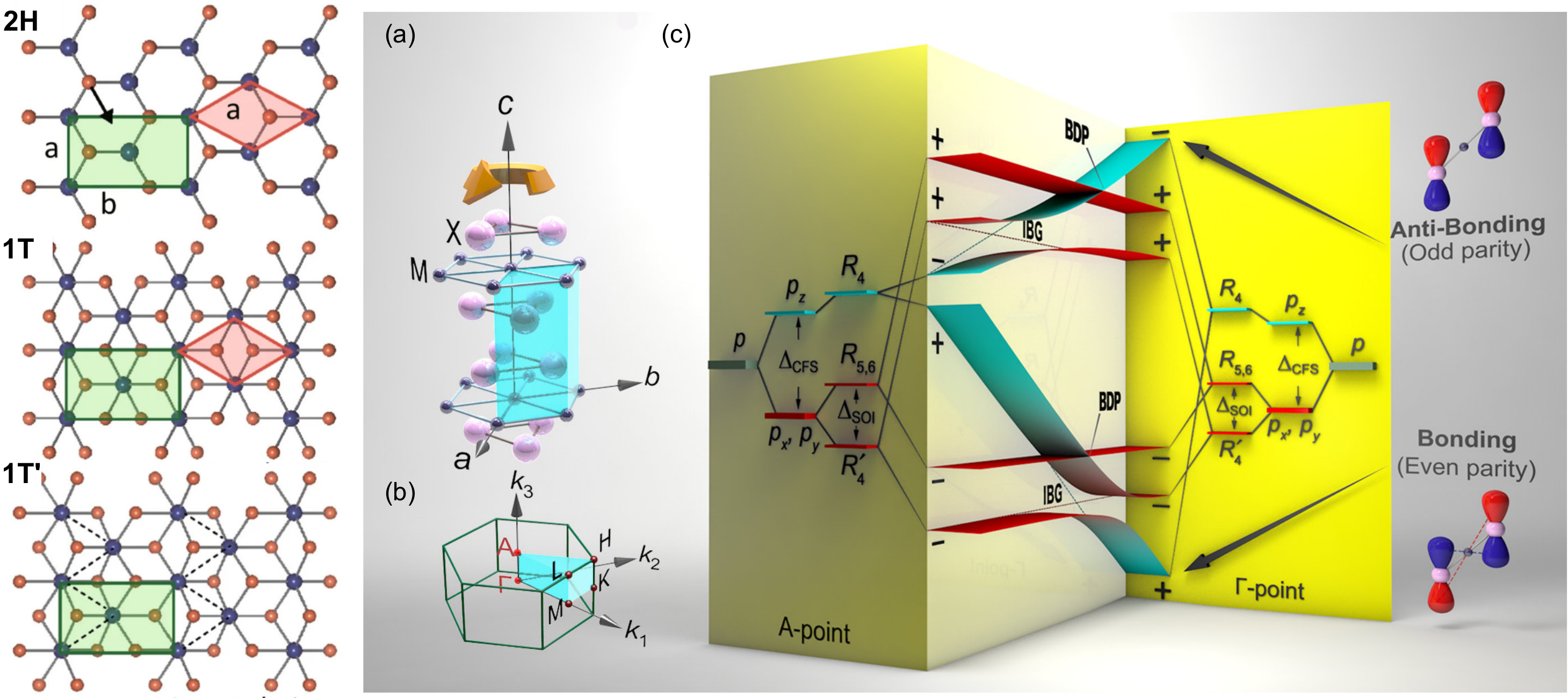}
	\caption{\label{TMDFig2}  Structural  and electronic properties of transition metal dichalcogenides. Left panel: Top-down crystal structures of single-layer transition metal dichalcogenides with 2H, 1T and 1T' (monoclinic) structures from high to low. Primitive unit cells are indicated with orange shaded areas. 1 $\times \sqrt{3}$ supercells, from which the primitive unit cell of the 1T' structure is derived, are indicated with green shaded areas. Panels taken directly from~\cite{chen_diverse_2021} (a-c) Mechanism underpinning the formation of bulk Dirac points and inverted band gaps in 1T structured TMDs. Crystal structure and Brilluoin zone for a 1T TMD, with the $C_{3\nu}$ rotational axis, and high symmetry momenta highlighted. (c) Effect of bonding-antibonding (B-AB) splitting, crystal field splitting and the spin-orbit interaction on the $p$-orbital manifold. Resulting bulk Dirac points (BDPs) and inverted band gaps (IBGs) are labelled. Taken from~\cite{clark_general_2019}.}
\end{figure*}

No material family better exemplifies the plethora of topological phases only recently verified to exist in previously well-studied compounds than the transition metal dichalcogenides (TMDs), a materials family exhibiting numerous examples of non-trivial band topology, bulk Dirac points and Weyl points, often simultaneously, and alongside many other desirable properties for next-generation device fabrication. The TMDs are a family of $>$30 stable compounds with the formula MX$_2$, where M is a transition metal and X $\in \{$S,Se,Te$\}$. Each of these van der Waals layered materials stabilizes in one or more of the following structures;  1T (P3$\overline{\text{m}}$1), 2H (P63/mmc), 3R (R3m) or in one of several distorted trigonal structures (triclinic ($\overline{P}1$), monoclinic ($\beta$, P2$_1$/m) and orhorhombic ($\gamma$, Pmn2$_1$)), here collectively referred to as 1T'. A subset of these is schematicized on the left hand side of Fig.~\ref{TMDFig2}~\cite{chen_diverse_2021}. 

The thermodynamically favoured structural phase of a TMD defines the crystal field splitting of the transition metal $d$-orbital manifold, which together with the electron filling factor, determines many of the ground state electronic properties of the TMDs~\cite{chhowalla_chemistry_2013}. For 2H (and 3R) structured TMDs ($D_{3h}$), the $d$-orbital manifold is split into three, energetically separated subsets ($a_1$, $e$ and $e'$ from low energy to high) located between the bonding and antibonding chalcogen $p$-orbital  manfolds and are found predominantly in  groups V ($d^1$ metals) and VI ($d^2$ semiconductors). 1T and 1T' structured TMDs instead have the $d$-orbital manifold split in two ($t_{2g}$ and $e_g$) with examples found in groups IV (1T $d^0$ semiconductors), V (1T $d^1$ metals), VI (1T' (monoclinic/orhorhombic) $d^2$ metals), VII (1T' (triclinic) $d^3$ metals), IX (1T $d^5$ metals) and X (1T $d^6$ (semi)metals). For the latter two cases, M-M bonds are preferentially formed over M-X bonds, placing the Fermi level within the anti-bonding $p$-orbital manifold. Both 1T and 2H $d^1$ metallic systems contain charge density instabilities, with the signatures of back-folded bands clearly seen with several spectroscopic techniques including ARPES~\cite{ritschel_orbital_2015, rossnagel_fermi_2005, wang_threedimensional_2021, feng_electronic_2018, chen_dimensional_2022, nakata_robust_2021, borisenko_two_2009, terashima_charge_2003}. These phases are often precursors to superconductivity at low temperatures, a property also shared by many of the metallic $d^5$ and $d^6$ TMDs. Perhaps most famously, the $d^2$ 2H semiconductors are the central focus of the burgeoning field of valleytronics owing to their graphene-like honeycomb lattice structure (Fig.\ref{TMDFig2}) and thus their layer-and momentum-locked Rashba-split valence band maxima (see e.g. the seminal photoluminescence studies Mak et al., 2012 and Xiao et al., 2012~\cite{mak_new_2012, xaio_coupled_2012}).

The topological properties of these materials only began to draw attention in 2015 with the prediction and subsequent verification of type-II Weyl nodes and Fermi arcs in the $d^2$ 1T' (orthorhombic, $\gamma$) structured tellurides, MoTe$_2$ and WTe$_2$ (see the review by Das et al.~\cite{das_electronic_2019}), establishing the TMD family as among the first hosts of type-II Weyl semimetals. 
In addition to this wide array of TMD properties deriving primarily from the transition metal $d$-orbital manifold, remarkably, the 1T and 2H TMDs were additionally found to host topological surface states and/or bulk Dirac points, the hallmarks of TIs and Dirac semimetals, respectively. In contrast to the Weyl phases of the 1T'-TMD subset, these topological properties stem entirely from within the chalcogen $p-$orbital manifold and are thus  largely insensitive to the choice of transition metal~\cite{bahramy_ubiquitous_2018}. Non-trivial band topology and/or bulk Dirac crossing points can therefore be found across the entire 1T and 2H TMD subclasses. 
Already, the topologically non-trivial TMDs have been shown to be of practical interest, with `giant’ or non-saturating magnetorestance observed in both TMD Dirac and Weyl semimetals (see e.g. Pavlosiuk et al., 2018 and Ali et al., 2014, respectively~\cite{pavlosiuk_gavanomagnetic_2018, 
ali_large_2014}), and tunable, high-efficiency THz detection realized in numerous group X Dirac semimetals (e.g. Guo et al., 2020, Cheng et al., 2023~\cite{cheng_giant_2023,  guo_anisotropic_2020}).

The type of the protected crossings available to each TMD compound (2D Dirac, 3D Dirac or Weyl) is determined only by whether an axis of rotational symmetry and/or an inversion center is present within the crystal structure. In the 1T and 2H structured TMDs, the $C_{3v}$-symmetric $k_z$ axis is capable of protecting any formed bulk Dirac points against perturbations including spin-orbit coupling \cite{yang_classifcation_2014}. By lifting inversion (or time-reversal) symmetry in such a way that the rotational axis is preserved, Dirac nodes could be transformed into Weyl points in the usual way. In contrast, the 1T' ($\gamma$) structured TMDs possess neither a sufficient rotational axis nor inversion symmetry. In this scenario, Weyl points can be instead formed at the intersection of electron- and hole-like states at low symmetry $k-$vectors, and are unrelated to the Dirac points in their 1T-structured counterparts. These two cases will be briefly overviewed below. 


\

\subsection{{\normalfont\normalsize\bfseries VIII. A. Dirac  cones in inversion symmetric TMDs}}

The mechanism underpinning the simultaneous formation of bulk Dirac points  and TSSs in bulk 1T and 2H structured TMDs has been extensively discussed in Refs~\cite{bahramy_ubiquitous_2018, clark_general_2019}, thus explaining the pre-existing observations of a topological surface state in PdTe$_2$~\cite{yan_identification_2015}, and bulk Dirac points in PtTe$_2$~\cite{yan_lorentz_2017} and PtSe$_2$~\cite{huang_type_2016}. This mechanism, applied to a 1T-TMD crystal structure, is schematisized in Fig.~\ref{TMDFig2}~\cite{clark_general_2019}.  In short, the bonding (B) and antibonding (AB) sets of the chalogen $p$-orbitals are each split into three doubly degenerate bands by the combined influence of the trigonal ($C_{3v}$) crystal field and spin-orbit interaction: The $p_z$ derived $R_4$ and the $p_{x,y}$-derived $R_{4}'$ and $R_{5,6}$, where the subscript denotes the irreducible representation. In van der Waals stacked systems, out-of-plane hopping across the van der Waals gap is in general larger for $p_z$ orbitals due to the larger spatial extent into the gap. The out-of-plane bandwidth (// $k_z$) of $p_z$-derived bands in momentum space is therefore larger than that of the pair of $p_{x,y}$-derived bands, making crossings between the six $p$-derived bands (3 for each AB and B sets) along $k_z$ highly likely. Around half of these crossings will be between bands of different irreducible representations, producing $C_{3\nu}$ lattice symmetry protected bulk Dirac points~\cite{yang_classifcation_2014}. The rest will hybridise to open up band gaps, which may be topologically non-trivial depending on the parity of the involved bands, which in real TMD systems typically correlates with the sign of group velocity of each band between the $\Gamma$ and A high-symmetry points. The result is a ladder of 3D and 2D Dirac cones, centered at the $\overline{\Gamma}$ point of the surface Brillouin zone and stacked along the energy axis. 

The bulk Dirac points produced by this mechanism occur at low symmetry $k$ points between $\Gamma$ and A and, as a result, typically exhibit anisotropic band dispersions in three dimensions, thus violating the Lorentz symmetry intrinsic to the Dirac fermion analogues in high energy physics. Instead, their dispersion is tilted, with the type of Dirac cone determined by whether the two branches of the cone have different (type-I) or the same (type-II) group velocity in analogy to the tilted Weyl cones of type-II Weyl semimetals. 

\begin{figure*}[t]
	\centering
	\includegraphics[width=\textwidth]{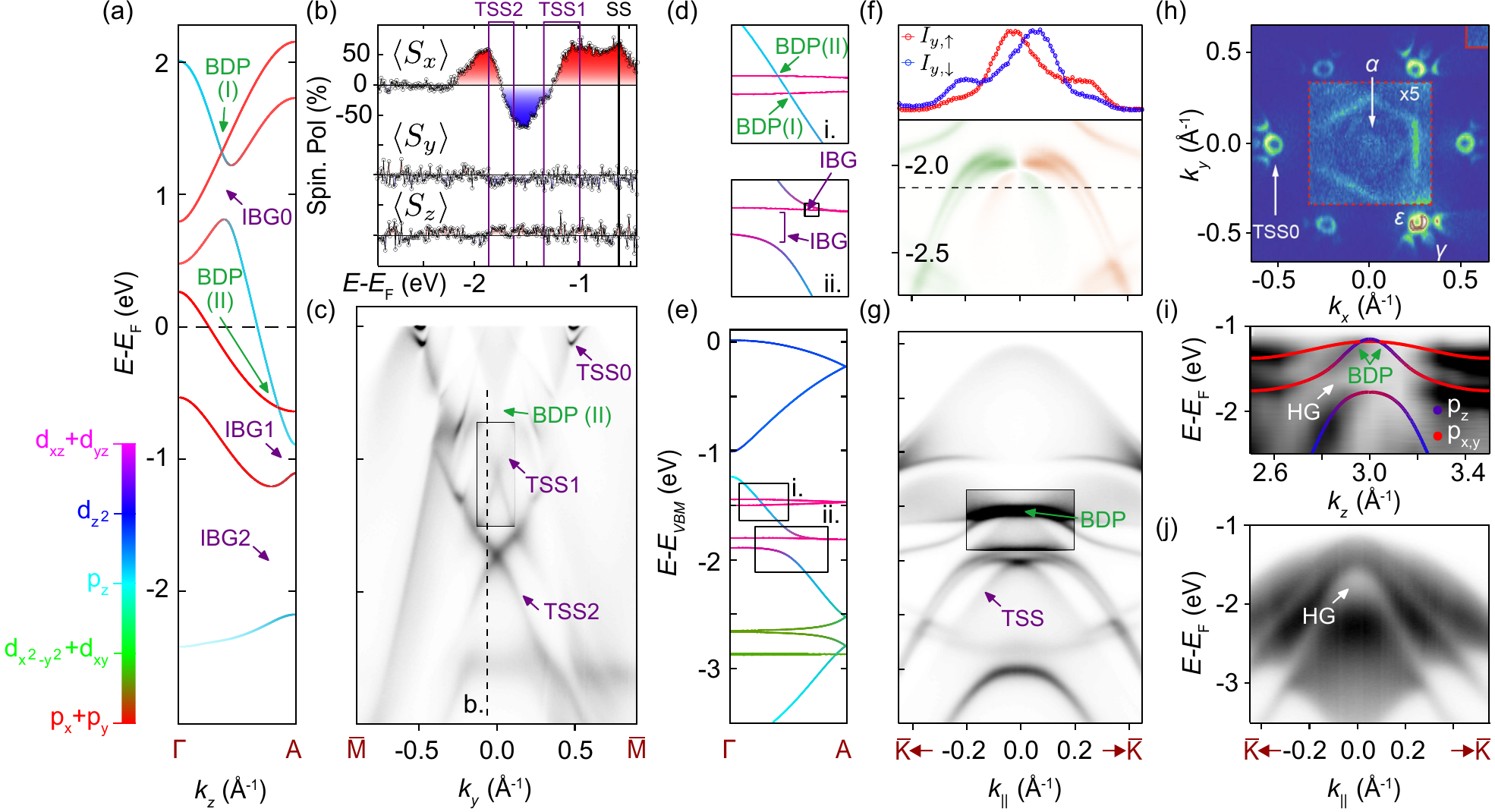}
	\caption{\label{TMDFig3} Topological  ladders in bulk transition metal dichalcogenides.  Compilation of ARPES and spin-resolved ARPES findings on (a-c) 1T-PdTe$_2$ ~\cite{clark_fermiology_2018, bahramy_ubiquitous_2018}, (d-g) 2H-WSe$_2$~\cite{bahramy_ubiquitous_2018}, (h) 1T-NiTe$_2$~\cite{mukherjee_fermi_2020} and (i-j) 1T-HfSe$_2$~\cite{clark_hidden_2022}. Orbitally-projected density functional calculations in (a, d-e) share the colour bar inset on the left, though the transition metal $d$-orbital contribution is neglected in (a). Calculations in (d) are zoom-ins of the indicated regions in (e). Circular dichroic spectrum in (f) is a zoom-in of the TSS region in (g) to better resolve the TSS. Highlighted (black box) region in (c) is performed with a higher photon energy to optimize matrix elements of TSS1. Highlighted regions in (g) and (h) have an increased image contrast only. Dashed lines in (c) and (f) indicate the energy ranges and approximate momentum positions for the spin-resolved energy and momentum distribution curves in (b) and (f) respectively (refer to original publications for details). Inverted band gaps (IBGs), trivial hybridsation gaps (HG), Type-I/II bulk Dirac points (BDPs (I/II)) and topological surface states (TSSs) are labelled throughout.}
\end{figure*}

Clear ARPES and density functional theory (DFT) evidence for topological ladders has been reported 
in the 1T systems of PdTe$_2$~\cite{yan_identification_2015, noh_experimental_2017, bahramy_ubiquitous_2018, clark_fermiology_2018, clark_general_2019}, PtTe$_2$~\cite{yan_lorentz_2017}, PtSe$_2$~\cite{huang_type_2016, bahramy_ubiquitous_2018}, NiTe$_2$~\cite{xu_topological_2018, ghosh_observation_2019, mukherjee_fermi_2020, nurmamat_bulk_2021}, CrTe$_2$~\cite{ou_zrte2_2022, li_tunable_2021}, HfSe$_2$~\cite{clark_hidden_2022}, CoTe$_2$~\cite{chakraborty_observation_2023} and IrTe$_2$~\cite{fei_band_2018, jiang_comprehensive_2020, nicholson_uniaxial_2021}.  The differences in the composition of the topological ladder depend largely on the starting energetics of the six doubly degenerate $p$-derived bands: a combination of the B-AB splitting (determined by the degree of orbital overlap), the crystal field strength, and the chalcogen spin-orbit coupling strength. For example, in PdTe$_2$ (Fig.~\ref{TMDFig3} (a-c)~\cite{bahramy_ubiquitous_2018, clark_fermiology_2018}), the bandwidth of the $p_z$-derived AB-$R_4$ band is large enough to cross through both the AB and B sets of $p_{x,y}$-derived bands as well as the oppositely dispersing B-$R_4$ band, thus creating a total of two bulk Dirac points and three topological surface states at $\overline{\Gamma}$. In contrast, in HfSe$_2$ (Fig.~\ref{TMDFig3}(i-j)~\cite{clark_hidden_2022}), the smaller chalcogen and larger van der Waals spacing leads to a much reduced  bandwidth of the $p_z$-derived and  no overlap between the B and AB derived $p$-orbital manifolds. The result is a topologically trivial hybridsation gap and a single type-II bulk Dirac cone formed along $\Gamma$-A within the $p$-orbital manifold, though it is particularly noteworthy that the pair of type-II bulk Dirac points (symmetric about A) comprises the valence band maximum of this compound~\cite{clark_hidden_2022}. 

Although typically buried deep within the valence bands due to the energetic positioning of the chalocgen manifold, similar physics was predicted by DFT in the 2H-structured TMDs NbSe$_2$, TaSe$_2$ and WSe$_2$~\cite{bahramy_ubiquitous_2018}. The 2H-structured variants have an extra consideration: by doubling the size of the unit cell along the $c$-axis, the number of distinct $p$-derived bands also doubles, thus providing twice the opportunities for bulk Dirac point and inverted band gap formation. For example, in 2H-WSe$_2$,  pairs of very closely spaced (in energy and momentum) inverted band gaps and bulk Dirac points (which are necessarily of opposite types due to opposite group velocity of the two halves of a back folded $p_{x,y}$-derived band) are predicted by density functional theory, with a topological surface state verified by spin-resolved ARPES~\cite{bahramy_ubiquitous_2018}, see Fig.~\ref{TMDFig3}(d-g)). The observable surface state has its dispersion heavily influenced by the restricting nature of the surrounding small $k_z$-projected band gap to produce a Rashba-like dispersion. This is a common phenomenon across these materials, with both trivial and topological surface states adopting highly usual dispersions due to the presence of only narrow channels left unexplored by the bulk electronic structure. This is particularly clear in PdTe$_2$~\cite{clark_fermiology_2018, clark_general_2019} and NiTe$_2$~\cite{mukherjee_fermi_2020}, where an inverted band gap above the Fermi level spawns a  topological state with a dispersion maximally warped by the dense $k_z$-projected bulk manifold, eventually crossing the Fermi level along  $\overline{\Gamma}$-$\overline{\mbox{M}}$ (Fig.~\ref{TMDFig2}(c,h)) with a highly complex in-plane spin texture~\cite{clark_fermiology_2018, mukherjee_fermi_2020}.


In conjunction with the Weyl semimetals within the TMD family, the addition of surface and bulk Dirac physics 
in the rotationally symmetric subclass establishes the TMDs as an ideal platform to study all types of topological band structure, and their interplay with other material specific properties. It is noteworthy that the generality of topological ladders to the rotationally symmetric material class as a whole is a direct consequence of the single-orbital manifold nature of the various band inversions contained within, and demonstrates an insensitivity of this non-trivial band topology to changes in spin-orbit coupling strength, $d$-orbital manifold overlap, and unit cell size. This is in sharp contrast to systems with band inversions between two distinct orbital manifolds, for example in the Bi$_2$Se$_3$ series, wherein the band inversions can be unwound by sufficiently altering the energetic separation of the Bi $p$ and Se $p$-orbital manifolds, for example, through reducing spin-orbit coupling.


Despite the robustness of the topological phases, the composition and energetics of the topological ladders in TMDs have been theoretically shown to be highly tunable through any perturbation altering the relative hopping parameters within the chalcogen sublattice, such as through alloying or through strain/compression~\cite{bahramy_ubiquitous_2018, huang_type_2016, xiao_manipulation_2017, clark_general_2019}. Of particular interest experimentally is the case of IrTe$_2$, which possesses a type-II bulk Dirac cone ~100 meV above the Fermi level as well as two inverted band gaps located at higher binding energy~\cite{bahramy_ubiquitous_2018, fang_structural_2013, nicholson_uniaxial_2021}. Tantalizingly, this compound also undergoes several temperature-driven structural phase transition from 280~K, eventually finding a 6 $\times$ 1 ground state (see Nicholson et al., 2021 and references within~\cite{nicholson_uniaxial_2021}) While the ground state typically has nanoscale domain sizes, the application of 0.1\% strain along the $a-$axis stabilises this 6 $\times$ 1 phase sufficiently to probe in isolation with ARPES. 

In principle, this change of symmetry should lift the protection of the bulk Dirac points due to the loss of $C_{3\nu}$ rotational symmetry, suggesting the possibility to exploit the structural transition as a temperature-controlled switch for bulk Dirac transport in this compound~\cite{yang_classifcation_2014}. In contrast, it was not only observed that some signatures of the bulk Dirac point seem to endure  this breaking of symmetry, but also that the energetics of the Dirac node fall below $E_F$ in the 6 $\times$ 1 phase and become the dominant interlayer transport channel~\cite{nicholson_uniaxial_2021}. This temperature-controlled Lifshitz transition is accompanied by the corresponding non-saturating linear magnetoresistance~\cite{zhang_superconductivity_2019}. Alternatively, a similar Lifshitz transition can be achieved in the high-temperature trigonal structure by doping with Pt~\cite{fei_band_2018, jiang_comprehensive_2020} which remains superconducting for Pt concentrations $<$ 0.4, a range that includes the composition where the bulk Dirac point is at $E_{\text{F}}$~\cite{fei_band_2018, jiang_comprehensive_2020}. These studies provide a proof of principle for tunable topological phases in TMDs in general, and  highlight the ever-growing phase space available for \textit{in situ} manipulation of electronic structures with ARPES. 

This single-orbital manifold origin responsible for both the robustness and tunability of these Dirac nodes is not limited to the TMD family, however. For example, prior predictions of type-II bulk Dirac points near the Fermi level in both the VAl$_3$~\cite{chang_type_2017} and YPd$_2$Sn~\cite{guo_type_2017} families can be understood within this same framework, with the bulk Dirac points protected by four-fold rotational symmetries in analogue to the three-fold symmetry protected Dirac crossings in the TMDs. More recently, PdTe, like PdTe$_2$, was found to host a series of $p$-orbital derived band crossings along the $\Gamma$-A direction of its P63/mmc type Brillouin zone by ARPES~\cite{yang_coexistance_2023}. While the presence of only one chalcogen per unit cell precludes any topologically non-trivial hybridisation gaps within the $p$-orbital manifold, a type-I bulk Dirac point is formed  within 100~meV of $E_{\text{F}}$. Crucially, unlike in the dichalcogenides, there is a possibility to realize a surface termination wherein pairs of bulk Dirac points formed along A-$\Gamma$-A do not project to the same point in the surface Brillouin zone, enabling the experimental characterisation of Fermi arcs spanning between bulk Dirac points. 



\begin{figure*}
	\centering
	\includegraphics[width=\textwidth]{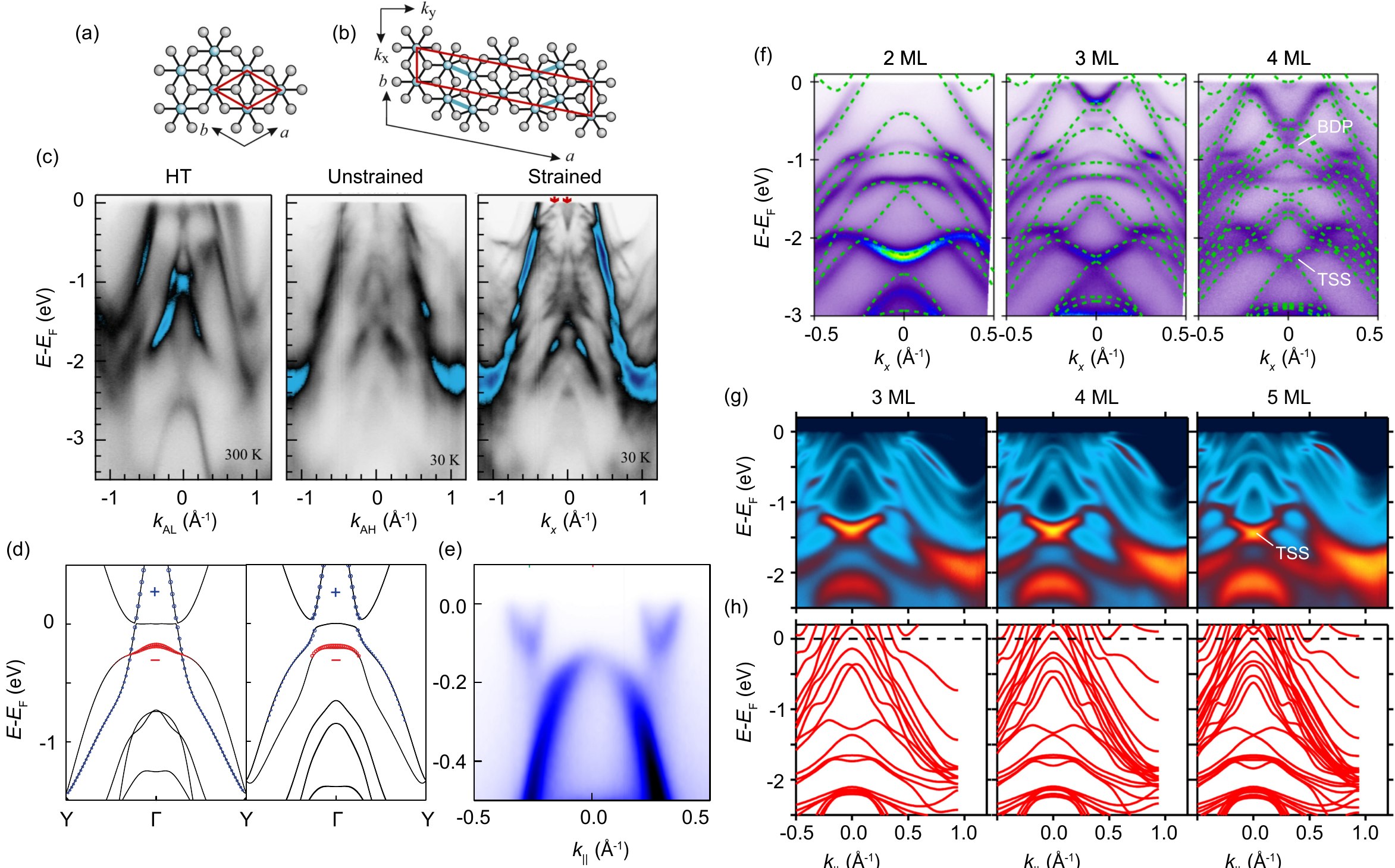}
	\caption{\label{TMDFig4} Tunable  topological phases in bulk and thin-film TMDs. (a-c) Temperature- and strain-dependent ARPES results on 1T-IrTe$_2$ adaped from Ref.~\cite{nicholson_uniaxial_2021}. (d-e) Band structure calculations and ARPES spectrum of (n-doped) monolayer 1T'-WTe$_2$ adapted from Ref.~\cite{tang_quantum_2017}. (f) Thickness-dependent ARPES of PtTe$_2$ thin films, adapted from Ref.~\cite{deng_crossover_2019}. (g-h) Thickness dependent ARPES and density functional theory calculations of NiTe$_2$ thin films, adapted from Ref.~\cite{hlevyack_dimensional_2021}. (a) High and (b) low temperature structures of 1T-IrTe$_2$. (c) High-temperature (HT), low-temperature unstained and strained ARPES spectra of 1T-IrTe$_2$. The application of ~0.1\% strain both drives a Lifshitz transition of the Dirac bands, and stabilizes the 6 $\times$ 1 ground state such that a single domain can be accessed. Red arrows indicate bands originating from the bulk Dirac cone in the high-temperature trigonal phase. Strain is applied along the $a$-axis.  (d) Band structure calculations of monolayer 1T'-WTe$_2$ without (left) and with (right) spin-orbit couling. The $\pm$ symbols indicate band parity at the $\Gamma$ point. (e) ARPES spectrum of 1T'-WTe$_2$ along the $\Gamma$-Y direction. The sample has been dosed with K to facilitate clear observation of the conduction bands. (f) ARPES spectra and overlaid band structure calculations for 2, 3 and 4 monolayer (ML) 1T-PtTe$_2$ (refer to original publication for further film thicknesses). Bulk Dirac points (BDP) and a topological surface states (TSS) are labelled. (g) ARPES spectra for 3, 4 and 5 ML 1T-NiTe$_2$ (refer to original publication for further film thicknesses). The TSS is labelled. (h) Corresponding density functional theory calculations for the spectra in (g).}
\end{figure*}

\subsection{{\normalfont\normalsize\bfseries VIII. B. Weyl phases in inversion asymmetric TMDs}}

\begin{figure*}
	\centering
	\includegraphics[width=\textwidth]{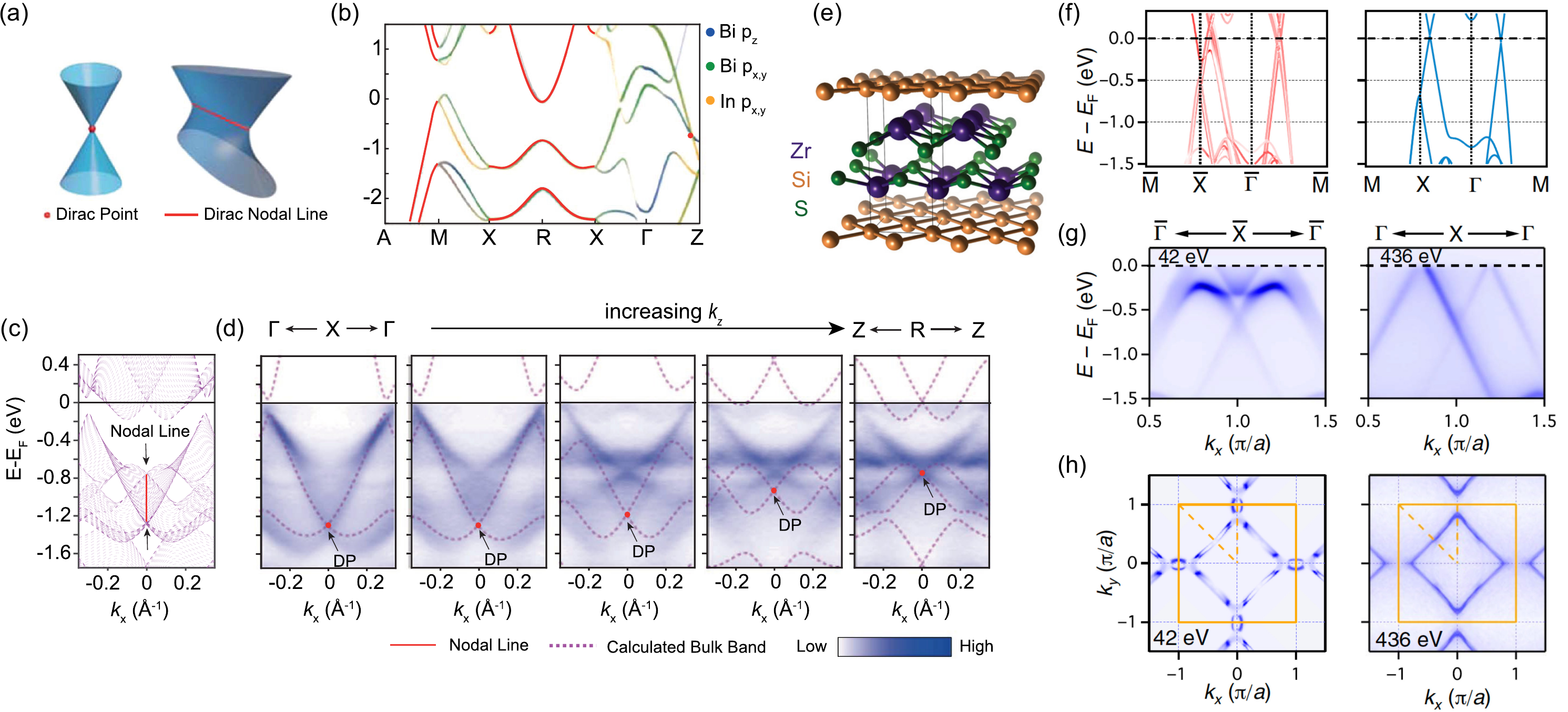}
	\caption{\label{TNLS}  Nodal lines and nodal surfaces in InBi and ZrSiS. Figures adapted from (a-d) Ref.~\cite{ekahana1_observation_2017}, (e) Ref.~\cite{pezzini_unconventional_2017} and (f-h) Ref.~\cite{fu_dirac_2019}. (a) Schematic of a Dirac point and Dirac nodal line. (b) Band structure calculation for InBi. Dirac points and Dirac nodal lines are overlaid with red markers/lines. Other colourings indicate orbital character. (c) $k_z$ integrated band structure calculations demonstrating the energy dispersion of the nodal line. (d) $k_z$-dependent datasets and overlaid band structure calculations demonstrating the continued presence of a protected Dirac crossing along the $k_z$ path. (e) Crystal structure of non-symmorphic ZrSiS. (f) Band structure calculations, (g) ARPES dispersions, and (h) ARPES Fermi surfaces of ZrSiS, for surface (left) and bulk (right) derived electronic structures. Yellow squares in (h) indicate the surface (bulk) Brillioin zones with the $\overline{M}$ (M) and $\overline{X}$ (X) points on the corners and edges, respectively.}
\end{figure*}

The first examples of topological phases in the TMDs were the cases of the distorted type-II Weyl semimetals $\gamma$-WTe$_2$ and $\gamma$-MoTe$_2$ discussed in Section~VI. Despite being the first topological phases predicted in the TMDs, the experimental verification proved much more challenging than that of the Dirac phases inherent to the inversion symmetric TMDs. 
Where the topological ladders in, for example, the group-X TMDs are ideal for direct observation in ARPES due to the large energy and momentum scales involved, the Weyl nodes in MoTe$_2$ and WTe$_2$ occur at low symmetry points between electron and hole pockets which disperse almost parallel to each other in the vicinity of the produced type-II Weyl points, thus leaving  energetically-narrow channels for Fermi arcs to disperse at all relevant $k$-points. Moreover, these crossings, protected by time-reversal symmetry, are located above the Fermi level which precludes imaging them directly with conventional ARPES~\cite{soluyanov_type_2015, das_electronic_2019}. This, combined with multiple surface terminations, the presence of trivial surface states alongside the predicted Fermi arcs~\cite{bruno_observation_2016} and a perceived extreme sensitivity of the Weyl nodes to strain and pressure~\cite{soluyanov_type_2015, bruno_observation_2016}, made an unambiguous classification of bulk MoTe$_2$ and WTe$_2$ and a complete mapping of Fermi arc spin textures difficult by purely spectroscopic means~\cite{das_electronic_2019}. 
Nevertheless, there is now significant spectroscopic evidence for Fermi arc surface states (and thus type-II Weyl points) in these compounds~\cite{wu_observation_2015, wang_observation_2016, sanchez-barriga_surface_2016, bruno_observation_2016, tamai_fermi_2016, jiang_signautre_2017, xu_evidence_2018, huang_spectroscopic_2016, wan_selective_2021}, with further evidence from transport experiments consistent with the presence of type-II Weyl points~\cite{hosur_recent_2013, udagawa_field_2016, li_evidence_2017}. In addition, both these compounds have been shown to be highly tunable. For example, the extremely large magnetoresistance in WTe$_2$, originating from a perfect compensation of charge carriers~\cite{ali_large_2014, alekseev_magnetoresistance_2012, das_electronic_2019}, is significantly amended following Lifshitz transitions that can be driven with temperature~\cite{wu_temperature_2015, pletikosic_electronic_2014} or pressure~\cite{kang_superconductivity_2015}. In MoTe$_2$, a structural transition from the orhorombic ($\gamma$) Weyl semimetal phase into the monoclinic ($\beta$) phase occurs above 240~K~\cite{xu_evidence_2018, he_dimensionality_2018}. This transition restores the inversion symmetry inherent to the undistorted 1T phase into the system, but not the three-fold rotational symmetry. This combination thus precludes the presence of the type-II Weyl nodes found in the $\gamma$ phase as well as the bulk Dirac nodes of the kind found across the 1T-TMDs~\cite{yang_classifcation_2014}. The compound remains interesting for its topological properties, however, as $\beta$-type 1T'-TMDs are theoretically predicted to contain higher order non-trivial band topology~\cite{tang_efficient_2019, wang_higher_2019}. This same $\gamma$-$\beta$ transition can be induced in 1T'-WTe$_2$ with ultrafast THz radiation~\cite{sie_ultrafast_2019}.





In addition to the Weyl physics in the 1T' TMDs, each bulk Dirac point in the 1T and 2H TMDs has the capacity to have its degeneracy reduced by breaking time-reversal or inversion symmetries to form Weyl nodes, so long as the underpinning $C_3$ symmetry remains in tact. While this has not yet been shown explicitly, this could be achieved by lifting time-reversal symmetry in a TMD system through the application of external magnetic fields or through the inclusion of magnetic dopants. Similarly, 1T-CrTe$_2$ may possess Weyl points intrinsically due to the combination of magnetic ground state (see for example Freitas et al., 2015, Zhang et al. for bulk and thin film variants~\cite{freitas_ferromagnetism_2015, zhang_room_2021}) and a chalcogen $p$-orbital manifold  which is likely to share the $k_z$-mediated band inversions of its sister compounds. Alternatively, inversion symmetry can be explicitly broken through the fabrication of artificial TMD structures with in-equivalent top and bottom chalcogen sub-layers, such as in the so-called `Janus' TMDs ( X$_1$-M-X$_2$, X$_1$ $\neq$ X$_2$), although this is not necessarily a sufficient criterion to form Weyl nodes, with Dirac (4-fold) and triple (3-fold) degenerate points maintained or formed in density functional theory studies of bulk Janus TMDs~\cite{xaio_inversion_2018, bahramy_ubiquitous_2018}. The creation of bulk Janus TMDs remains largely theoretical due to the logistics of such a precise sample growth~\cite{maghirang_predicting_2019, griffith_enhancing_2023}, though monolayer Janus TMDs in 1H, 1T and 1T' structures have been successfully synthesized~\cite{tang_2D_2022, li_recent_2018, zhang_janus_2017, trivendi_room_2020, zheng_janus_2022, gan_chemical_2022} but have yet to be measured by ARPES. 





\subsection{{\normalfont\normalsize\bfseries VIII. C. Thin Limit}}

While bulk TMDs offer a rich platform in which to study and induce a wide array of novel physics, next-generation electronic or spintronic devices harnessing TMD materials are likely to turn to their few-monolayer variants both for easier control through gating~\cite{NguyenNature19, lei_graphene_2022} and for optimized integration with other materials~\cite{ji_recent_2022, pham_2d_2022}. Fortunately, the van der Waals nature of TMD systems makes them ideal for epitaxial growth via molecular beam epitaxy (MBE)  on a wide array of substrates, including  graphene, which is both highly conductinve and largely inert (see e.g. the following ARPES studies:~\cite{zhang_room_2021, ou_zrte2_2022, coelho_charge_2019, feng_electronic_2018, biswas_ultrafast_2021, dreher_proximity_2021, watson_strong_2021, antonelli_orbital_2022}). Similarly, numerous methods of mechanically exfoliating monolayers from bulk single crystals are possible, rendering high quality films which can be integrated into devices, though the size of such films is typically on the order of 20~$\mu$m. The development of nano-ARPES (i.e. ARPES with focused light spots smaller than 1~$\mu$m) and angle-resolved photoelectron emission microscope (PEEM) beamlines worldwide has therefore proven crucial for the systematic characterisation of these exfolated TMD flakes and their devices~\cite{grubisic_insitu_2023, stansbury_ws2wse2_2021, pei_observation_2022, khalil_wse2mose2_2022, NguyenNature19, muzzio_visualizing_2021}, and for TMD thin films more generally which are often spatially inhomogeneous and suffer from rotational disorder. Moreover, MBE grown or exfoliated TMD flakes can be capped with amorphous Se and/or Te to prevent film degradation between growth chamber and ARPES ultra-high vaccuum environments, or often simply annealed \textit{in situ} without capping to remove surface contaminants.


For the 1T' structured TMDs, thinning  MoTe$_2$ down to the few layer limit (less than 12 nm) has been found to remove the structural transition to the inversion-symmetric monoclinic phase, thus providing access to the type-II Weyl nodes, usually found only within the low-temperature inversion asymmetric orthorhombic ($\gamma$) phase, at room temperature~\cite{he_dimensionality_2018}. In the single-layer limit, both MoTe$_2$ and WTe$_2$ become inversion symmetric (though inversion symmetry is removed with a small perpendicular electric field) since the $\beta$ and $\gamma$ phases differ only by their stacking pattern along the $c-$axis, and are therefore equivalent in the single layer~\cite{rhodes_enhanced_2021}. While a 2D analogue of Weyl nodes may then not be expected, monolayer 1T' WTe$_2$ has been shown to be an intrinsic two-dimensional topological insulator via numerous experimental methods including ARPES (Fig.~\ref{TMDFig4}(a)), with inverted band gaps formed within the $d$-orbital manifold near the Fermi level generating topologically protected, conductive spin-polarised edge states around the perimeter of the semiconducting film~\cite{tang_quantum_2017, fei_edge_2017, shi_imaging_2017, zhao_determination_2021}. 


For the 1T and 2H-structured TMDs, one would not  expect topological ladders of bulk and surface Dirac nodes to survive the thinning-down to a single monolayer. Their formation, along the $k_z$ axis, relies the on the intra- and  inter-layer hopping channels perpendicular the surface plane, which are strongly reduced in the few layer limit. Indeed, no pristine monolayer TMDs have been found to host any Dirac nodes. For example, in its bulk 1T-PtTe$_2$ is metallic and hosts two bulk Dirac cones and three inverted band gaps at $\overline{\Gamma}$~\cite{yan_lorentz_2017}. In its monolayer, it is a topologically trivial semimetal~\cite{deng_crossover_2019}. However, the topological physics reappears abruptly as soon as 4 monolayers, with clear TSS and `bulk' Dirac states observed by ARPES in MBE grown films (Fig.~\ref{TMDFig4}(f)~\cite{deng_crossover_2019}. A similar phenomenon was observed in semimetallic NiTe$_2$ films, where gapless TSS are found in five-layer samples, alongside conical bands deriving from the (near-$E_\text{F}$) type-II bulk Dirac cone in the bulk limit (Fig.~\ref{TMDFig4}(g-h)~\cite{hlevyack_dimensional_2021}). 
These studies suggest that the topological crossing points are not gradually unwound with reduced thickness, but rather that the hierarchy of band inversions 
is left in-tact along a discretized $k_z$ axis. The layer number needs only be sufficient to (a) access $k_z$ sub-bands where bulk Dirac points form, and/or (b) access sub-bands on the non-trivial side of the $k_z$-mediated topological phase transitions which generate the surface Dirac nodes. As with all TIs, the film needs also to be sufficiently thick to avoid wavefunction interference of topological surface states on opposing surfaces. Topological phases may also be induced in monolayer 1T-TMDs via chemical substitution. For example, in monolayer 1T-PtSe$_2$ a 2D topological insulating phase, unrelated to non-trivial band topology of the bulk, has been theoretically shown to be induced through Se-Hg substitution (with one non-trivial alloy naturally occurring), with the resulting topologically non-trivial edge states crossing the Fermi level~\cite{crastodilima_toplogical_2023}.





Altogether, the presence of both bulk Dirac nodes and non-trivial topology in the few-layer TMD limit 
opens avenues to incorporate topological physics in thin film heterostructures. For example, designer topological ladders, with selective control over the numbers, types and energetics of Dirac nodes, may be engineered in 1T and 2H TMD heterostructures with a careful choice of TMD on a layer-by-layer basis. Such heterostructures could also open routes to drive correlated topological phases not possible in the bulk systems. In analogy to the induction of Mott gaps and superconductivity in bilayer graphene through the  delicate control of the relative twist angle between the two monolayers~\cite{cao_unconventional_2018}, it has been shown how correlated phases are achievable in both twisted heterobilayer and homobilayer TMDs. In short, the moir\'e superlattice potential defined by the lattice mismatch between constituent layers enables hybridization channels between states that were previously well separated in momentum space, due to extreme back-folding into the reconstructed `mini' Brillouin zone~\cite{cao_unconventional_2018}. This produces non-dispersive `flat' bands from which correlated phases can emerge with the right band filling. Such flat bands have already been observed spectroscopically for numerous TMD heterostructures~\cite{stansbury_ws2wse2_2021, pei_observation_2022, li_twistedblwse2_2021}, with a correlated insulator phase experimentally found at half-band filling~\cite{wang_correlated_2020}. While such findings already allude to technological applications such as the Mott field effect transistor, the coexistance of non-trivial band topology and these engineered correlated phases is yet to be explored, but is in reach, as demonstrated by the verification of topological states in monolayer WTe$_2$~\cite{tang_quantum_2017}, monolayer PtHg$_x$Se$_{2-x}$~\cite{crastodilima_toplogical_2023} and 4-5 layer PtTe$_2$~\cite{deng_crossover_2019} and NiTe$_2$~\cite{hlevyack_dimensional_2021}.

\subsection{{\normalfont\normalsize\bfseries IX. Topological nodal-line semimetals}}

\begin{figure*}[t]
	\centering
	\includegraphics[width=\textwidth]{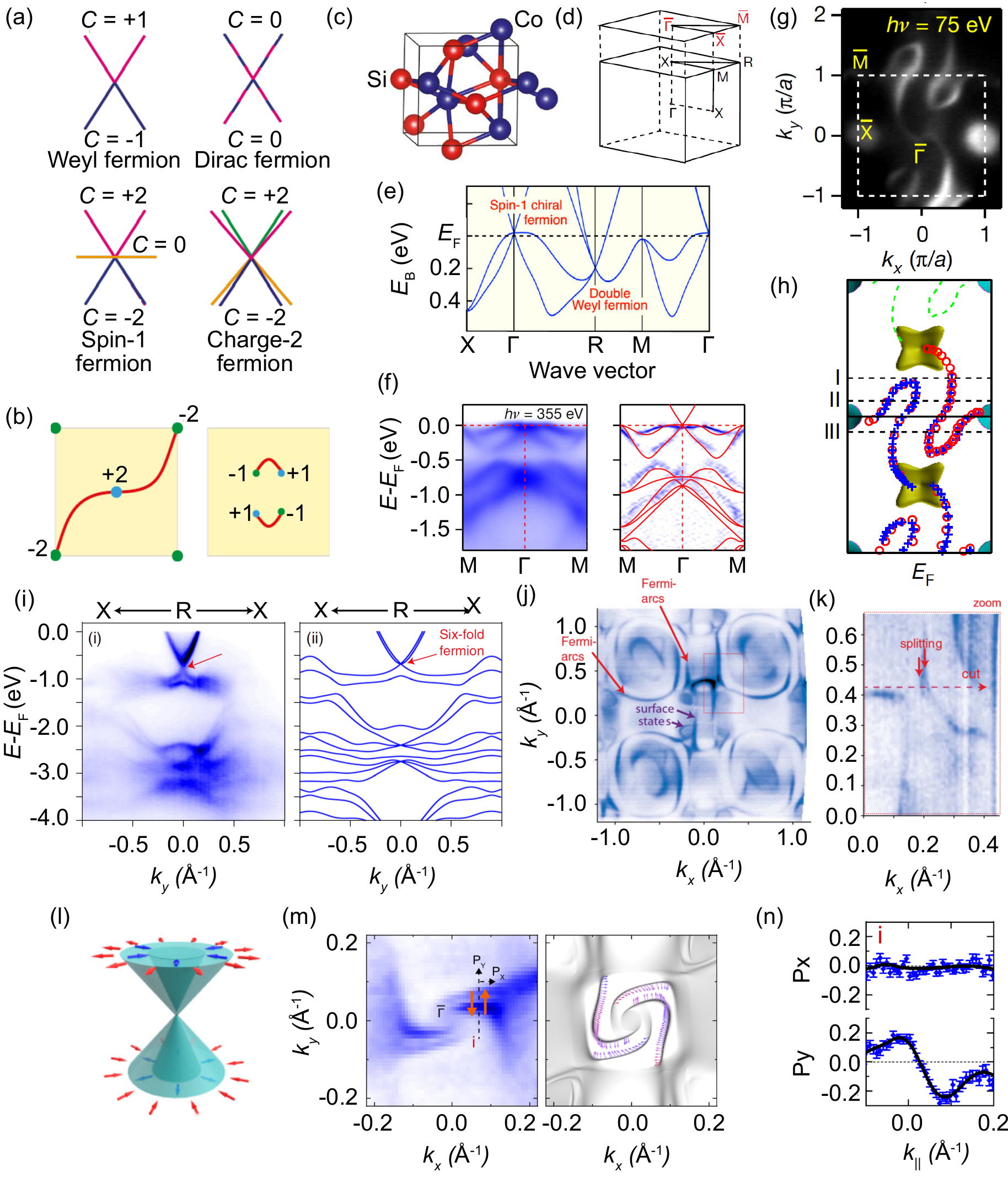}
	\caption{\label{FigChiral}  
Multifold fermions in topological chiral semimetals CoSi, PtAl, PdGa, and PtGa. (a) Examples of multifold fermions as compared to Weyl and Dirac fermions. (b) Since they are located at time-reversal invariant momenta, the surface Fermi arcs extend across the complete surface Brillouin zone (left), in contrast to the case of Weyl semimetals (right). (c-h) CoSi with (c) crystal structure, (d) bulk and surface Brillouin zone, and (e) band structure. (f) Bulk-sensitive ARPES data (355 eV) of the spin-1 chiral fermion. (g,h) Surface fermi arcs from data at two different photon energies (75 and 110 eV) superimposed on the calculated bulk Fermi surface. 
(i) In PtAl the large spin-orbit splitting of $\sim100$meV can be resolved. (j,k) Likewise, the surface Fermi arc of PdGa is split. (l) Spin texture of a multifold fermion. (m,n) Measurements of the spin texture of the surface Fermi arc of PtGa allow conclusions on parallel spin-momentum locking of the bulk multifold fermion. Panels (a,b,d-h) from \cite{RaoNature19}, (c) from \cite{TakanePRL19}, (i) from \cite{SchroeterNP19}, (j,k) from \cite{SchroeterScience20}, (l-n) from \cite{Krieger22}.
 }
\end{figure*}

\begin{figure*}[t]
	\centering
	\includegraphics[width=\textwidth]{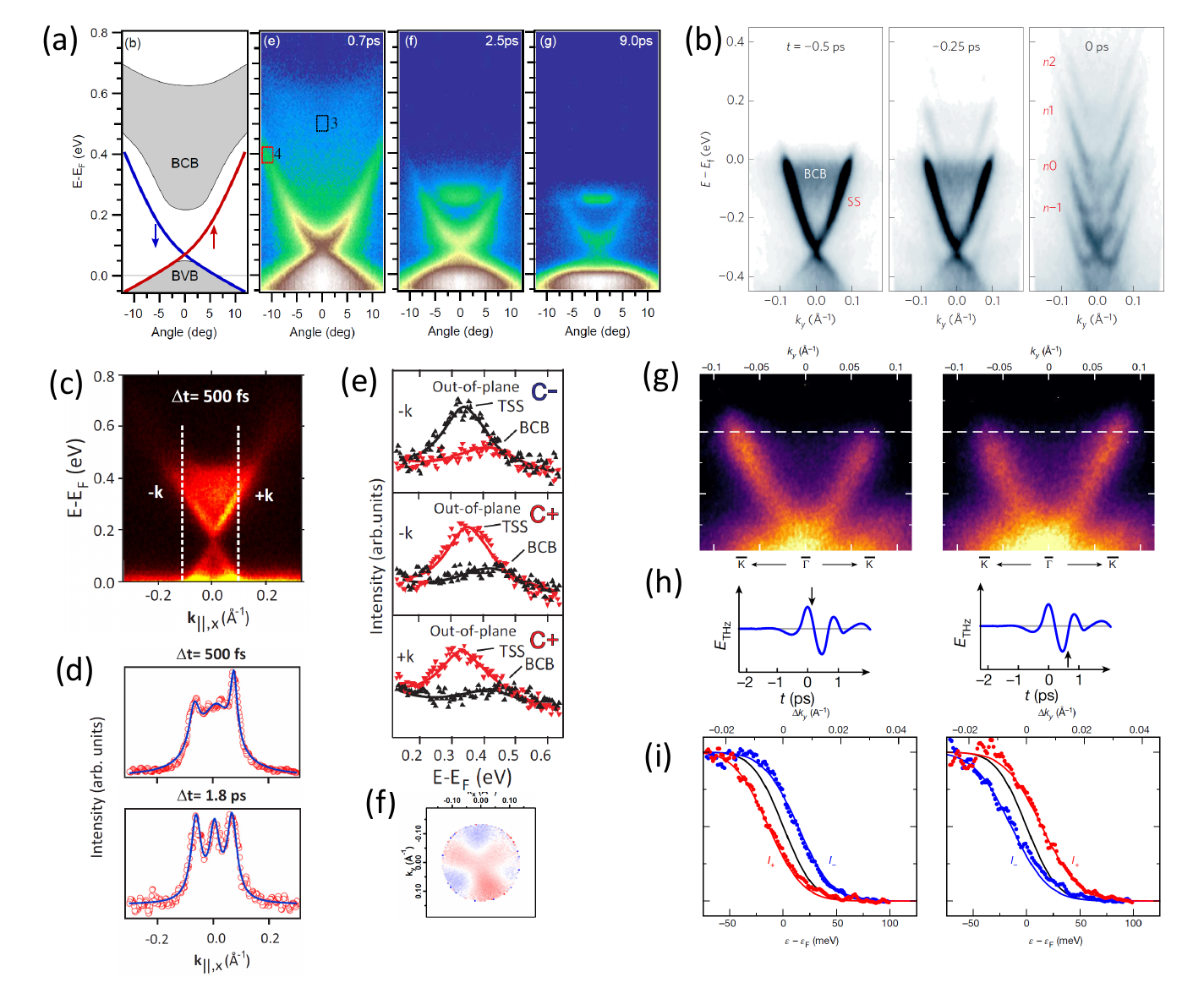}
	\caption{\label{FigUltrafast}  
Ultrafast pump-probe spectroscopy from Bi$_2$Se$_3$ and Sb$_2$Te$_3$. 
(a) Ultrafast populated unoccupied bulk conduction band states and the upper Dirac cone of p-type Bi$_2$Se$_3$ decay with a time constant which increases towards the Dirac point due to a bottleneck effect.
(b) Bloch-Floquet states in Bi$_2$Se$_3$ due to coherent interaction of photons and electrons. Intensity of sidebands is maximum where pump and probe pulse maximally overlap ($t=0$ ps).  
(c,e) Circularly polarized pump pulses create in Sb$_2$Te$_3$ a transient out-of-plane spin polarization in the Dirac cone
determined by the light helicity, here shown for $t=500$ fs.   (c,d) Circular dichroism reveals that more electrons with momentum along $+${\bf k} can be excited than along 
$-${\bf k}. (f) This asymmetry in the Dirac cone has for Bi$_2$Se$_3$ a 3-fold symmetry excluding a net effect on the (spin-polarized) photocurrent. (g-i) Acceleration of electrons in the TSS of Bi$_2$Te$_3$ by a THz pulse. The effect reverses with the sign of the electric field, shown in panel (h). (i) Blue (red) denotes the branch of the Dirac cone for negative (positive) momenta. 
Panel (a) taken from \cite{Sob12}, panel (b) from \cite{MahmoodNP16}, panels (c-e) from   \cite{San16b}, panel (f) from  \cite{Ket18}, panels (g-i) from  \cite{ReimannNature18}.}
\end{figure*}

 In analogy to the zero-dimensional nodes discussed in the previous Sections, it is possible to produce one dimensional nodal lines along complete $k$-paths through a Brillouin zone (schematisized in Fig.~\ref{TNLS}(a)). The hosting compounds, Dirac nodal line semimetals, often require additional crystalline symmetries with respect to Dirac and Weyl semimetals to ensure the protection of Dirac degeneracies over extended $k$-paths, at least in the absence of spin-orbit coupling. These nodal lines can take the form of nodal rings, chains links and knots~\cite{KimWieder15,FangNodal15,Bzdusek16,bi_nodal_2017,fu_dirac_2019} and, while the nodal line stays ungapped (neglecting spin-orbit coupling) over the full momentum range, the Dirac node is free to move in energy.
By now, many examples of nodal line semimetals have been theoretically predicted and/or experimentally confirmed with ARPES, including Mg$_3$Bi$_2$~\cite{chang_realization_2018}, which possesses nodal lines only in the absence of spin-orbit coupling due to the presence of only time-reversal and inversion symmetries, and PbTaSe$_2$~\cite{Bian2016}, Cu$_3$PdN~\cite{yu_topological_2015}, Ca$_3$P$_2$~\cite{xe_new_2015}, the Ca$_3$P family~\cite{xu_topological_2017} and  EuAs$_3$~\cite{cheng_magnetism_2021} which host nodal lines protected in the absence of spin-orbit coupling due to reflection/mirror symmetries, and occasionally in the presence of spin-orbit coupling (TaPbSe$_2$) depending on the Brillouin zone path where the nodal line occurs and the presence of inversion symmetry. It is noteworthy that the addition of spin-orbit coupling transforms a subset of these compounds into Dirac semimetals with lattice symmetry protected bulk Dirac points~\cite{yu_topological_2015}. 

There are instances where the nodal lines are resilient to spin-orbit coupling gapping along extended $k-$paths however. These cases employ non-symmorphic symmetry: A combination of a fractional lattice symmetry and either a glide plane or screw axis~\cite{wu_nonsymorphic_2022, wang_hourglass_2016, yang_symmetry_2018}. Non-symmorphic compounds with nodal lines include InBi (Fig.~\ref{TNLS}(b-d)~\cite{ekahana1_observation_2017}, 
Pt$\{$Sn,Pb$\}$$_4$~\cite{WuPtSn4, wu_nonsymorphic_2022},  
and 
(Ta,Nb)$_3$SiTe$_6$~\cite{li_nonsymmorphic_2018}. Perhaps most extensively studied, however, is the Matlockite (P4/nmm) materials families M$_1$M$_2$X, where M$_1$ $\in$ \{Zr, Hf, La, Na\}, M$_2$$\in$ \{Sb, Ge, Si\}, X$\in$ \{S, Se, Te\} which have been extensively characterised by ARPES~\cite{SchoopNC16,NeupanePRB16,ChenHfSiS17, topp_surface_2017, chen_dirac_2017, hosan_tunability_2017, wang_evidence_2016, yen_dirac_2021, chen_visualizing_2019, fu_dirac_2019, wang_spectroscopic_2021, lou_emergence_2016}. These compounds contain zero-dimensional Dirac nodes protected by the non-symmorphic symmetry itself, which here results from a glide-mirror symmetry with respect to the Si layer in ZrSiS, for example  (Fig.~\ref{TNLS}(e)), and one dimensional nodal lines protected by the combination of inversion and time-reversal symmetries~\cite{SchoopNC16}. The latter continuously connect Dirac crossings found along the high-symmetry X-$\Gamma$ and $\Gamma$-M directions to form a line of Dirac degeneracies along low-symmetry paths near the Fermi level (Fig.~\ref{TNLS}(f)). The nodal line itself is not protected against spin-orbit coupling (unlike the non-symmorphic symmetry protected Dirac nodes at higher binding energies), but in the case of ZrSiS the small spin-orbit coupling strength renders only millivolt scale bandgaps~\cite{fu_dirac_2019, yang_anisotropic_2021}. Note that this is different to the case of, e.g., InBi (Fig.~\ref{TNLS}(c-d)) where the non-symmorphic symmetry does fully protect the nodal lines even in the presence of strong spin-orbit coupling.
More recently, it was predicted~\cite{topp_surface_2017} and subsequently confirmed~\cite{fu_dirac_2019} that the nodal lines in ZrSiS form part of a nodal surface on the $k_z=0$ and $\pi$ contours of the Brillouin zone to produce a square-shaped bulk Fermi surface composed only of the near-Dirac degeneracies~(Fig.~\ref{TNLS}(g-h)).

It is noteworthy that these systems host particularly rich surface electronic structures where the non-symmorphic symmetries are broken. Indeed, in the ZrSiS family, the natural cleavage plane is between the Si layers, resulting in surface-derived ‘floating’  pockets  at the $\overline{X}$ point of the surface Brillouin zone, which accompany the Dirac nodal surface~\cite{SchoopNC16, topp_surface_2017, fu_dirac_2019, yang_anisotropic_2021}, see Fig.~\ref{TNLS}(f-h). These surface electronic structures can have rich properties of their own, including the formation of mirror-symmetry protected crossings between pairs of Rashba-split surface states, which behave as 2D analogues of Weyl points~\cite{markovic_weyl_2019}. 




\subsection{{\normalfont\normalsize\bfseries X. Topological chiral semimetals}}

Chiral crystals   are described by the 65 Sohncke space groups   which contain only symmetry operations of the first kind (rotations and screwings), that is, they lack   inversion, mirror, and roto-inversion symmetry. These   have been suggested to host multifold fermions 
which go beyond the  Weyl, Dirac and Majorana fermions of quantum field theory and of which the Weyl fermion is itself chiral. The discovered multiple degeneracies  include  three-, six-, and eight-band crossings stabilized by space group symmetries with spin-orbit coupling and time-reversal symmetry \cite{BradlynScience16}.   

The low-energy effective Hamiltonian describing, for example,  the 3-fold crossing is a fermionic   generalization of the Weyl hamiltonian $H={\bf k} \cdot {\bf S}$, where  {\bf k} ís the momentum and
 the matrices $S_i$ are the generators of the rotation group $SO(3)$ in the spin-1 representation. 
The  bands of the multifold fermions bear a higher Chern number,  $\pm2$ in the case of the 3-fold crossing, and also host a larger number of topologically protected surface Fermi arcs. The 3-fold fermions are predicted to show an anomalous negative magnetoresistance and a chiral anomaly different from that of single or double Weyl points. At large magnetic fields the Landau level spectrum displays two chiral modes with unusual momentum dependence \cite{BradlynScience16}.  

3-fold crossings have been observed before in MoP \cite{LvNature17} and WC  \cite{MaNatPhys18}. Those, however, result from band inversion, and do not lead to defined Chern numbers. The multifold fermions, instead, do not require band inversion and occur at time-reversal invariant momenta. In contrast to Weyl fermions they cannot move and annihilate and are therefore more robust as long as the relevant crystal symmetries are preserved. Single crossings of  spin-orbit coupled bands  at time-reversal invariant momenta have been named Kramers-Weyl fermions \cite{ChangGNM18}.

In the presence of time-reversal symmetry, multifold fermions occur for several space groups. Among these groups $P2_1 3$ (number 198) corresponds to a simple-cubic lattice where the 6-fold degeneracy occurs at the R point (corner of the Brillouin zone) \cite{BradlynScience16}. Space group 198 has been of particular interest because simple binary crystals are available such as transition metal monosilicides, monogermanides, monoaluminides, and monostannides.  Specific predictions have been made, for example, for silicides \cite{TangPRL17,ChangGPRL17,Pshenay18}.

ARPES data have simultaneously been reported for {\braun PtAl} \cite{SchroeterNP19}, CoSi 
\cite{RaoNature19,TakanePRL19,SanchezDSNature19}, and RhSi \cite{SanchezDSNature19}.  
The associated large Chern numbers lead to  multiple   Fermi arcs. The chirality of the crystal structure can be deduced from the Fermi arc. 
Fig.~\ref{FigChiral}(a) shows the comparison of the Weyl and Dirac fermion and multiple fermions: the chiral spin-1 fermion and the chiral double Weyl or charge-2 fermion.  Fermi arcs span the surface projection of the high symmetry points at which the multifermions occur and, therefore, can span the whole surface Brillouin zone, depending on the surface. They are hence typically an order of magnitude longer than in Weyl semimetals, shown as sketch in Fig.~\ref{FigChiral}(b), and this is observed in ARPES
\cite{SchroeterNP19,RaoNature19,TakanePRL19,SanchezDSNature19,SanchezDSNature19}. For CoSi, a spin-1 fermion occurs at $\Gamma$ [probed at soft-x-ray energies in Fig.~\ref{FigChiral}(e)] and double Weyl fermions at the R points. The surface Fermi arcs span the whole (100) surface Brillouin zone from
$\overline{\Gamma}$ to $\overline{M}$, see Fig.~\ref{FigChiral}(g,h).  There ARPES signature is independent of the photon energy due to their two-dimensionality (blue crosses 75 eV and red circles 110 eV in Fig.~\ref{FigChiral}(h)). 

For chiral semimetals with light elements the spin-orbit coupling is small and not resolvable in ARPES. For PtAl, PdGa, and PtGa the spin-orbit coupling is, however, high and the splitting can be resolved in the bulk crossing of PtAl in Fig.~\ref{FigChiral}(i) \cite{SchroeterNP19} and in the Fermi surface arc of PdGa in Fig.~\ref{FigChiral}(j,k) \cite{SchroeterScience20}
. The splitting of the surface Fermi arc implies that four arcs connect the projections of the multifold fermions, which is consistent with them having a large Chern number with magnetude 4. 

Topological chiral semimetals display a peculiar spin texture. We note that in Rashba systems the presence of mirror symmetries render the spin perpendicular to the mirror plane for states with wave vector in the mirror plane because an in-plane spin reverses under mirror reflection. A radial spin texture, that is, parallel spin-momentum locking, has been predicted for chiral elemental semiconductors Te and Se  \cite{HirayamaPRL15}. The Weyl nodes have been identified in Te by ARPES \cite{Nakayama17} and the radial spin texture was observed near the Fermi energy by spin-resolved ARPES for chiral Te \cite{SakanoPRL20,GattiPRL20} including its reversal for crystals with opposite chirality \cite{SakanoPRL20}. 
Gyrotropic effects, i. e., those that reverse with the crystal chirality, have been predicted for Te. 
A current along the trigonal axis induces a  small parallel magnetization, which can   be detected by  Faraday rotation of transmitted light \cite{Tsirkin18}. 

Te does not exhibit topological semimetal behavior and the radial spin texture is not expected to be maintained in a large range of momentum space. However, theoretical studies have predicted the presence of a truly radial spin texture (Fig.~\ref{FigChiral}(l)) and an isotropic parallel spin-momentum locking in  chiral topological semimetals with single- and multifold band crossings (\cite{ChangGNM18,LinMao22}). Spin-resolved ARPES investigations have been conducted on PtGa  revealing parallel spin-momentum locking for the multifold fermions at  $\Gamma$ and R. This was possibe through 
the spin texture of its topological Fermi arc surface states which points orthogonal to the Fermi surface contour close to the projection of the bulk multifold fermion \cite{Krieger22}. 
This is seen in Fig.~\ref{FigChiral}(l-n). Additionally, it has been demonstrated that the direction of spin polarization reverses with the chirality of the crystal \cite{Krieger22}.

\subsection{{\normalfont\normalsize\bfseries XI. Correlated topological insulators}}

The topological materials discussed so far  are classified based on their single-particle band structure and do not rely on electron correlation. Nonetheless, various research directions explore the connection between topology and strong correlation. One notable limitation of TIs identified thus far is the relatively small size of their inverted bulk band gaps, which restricts their practical use in semiconductor electronics. However, the Coulomb interaction can generate or augment band gaps to much greater magnitudes. In that case their gaps are centered about the Fermi level and do not require a certain stoichiometry such as in (Bi,Sb)$_2$(Te,Se)$_3$ compounds to become bulk insulating \cite{Shitade09}.

Generally the electron correlation is stronger among $d$ and $f$ electrons than for $s$ and $p$ electrons. In 5$d$ transition metal oxides, both the spin-orbit interaction and the electron correlation are strong and of the same order of magnitude. According to a tight-binding model, Na$_2$IrO$_3$ has been predicted to be a quantum spin-Hall insulator, which is  a 2D TI \cite{DzeroGalitski13}. It is important to mention that there are strongly correlated topological phases of matter that have no analogue in band theory and hence do not lend themselves to an investigation of their nontrivial topology by ARPES. This applies to the topological Mott insulator \cite{Raghu08,Pesin10} and extends to frustrated quantum magnetism \cite{BalentsNature10,Reuther12}. 

A prediction  directly related to the band structure and hence relevant for ARPES applies however to SmB$_6$ which has been predicted as topological Kondo insulator \cite{Dzero10,Alexandrov13,Takimoto11,LuF13}. The band gap in a Kondo insulator is caused by hybridization of itinerant conduction electrons, here $d$ electrons, with localized electrons, here $f$ electrons. Importantly, Dzero {\it et al}. realized that due to the odd parity of $f$ states and even parity of $d$ states, the precondition for band inversion is fulfilled \cite{Dzero10}. Band calculations predicted SmB$_6$ as strong TI with an odd number of Dirac cones on the (100) cleavage plane \cite{Alexandrov13,Takimoto11,LuF13}. 

SmB$_6$ was the first established mixed valent compound and Kondo insulator and the predictions as TI prompted further experimental efforts (see the review by Allen \cite{Allen16}). Surface transport experiments confirmed subsequently that at low temperatures only the surface remains metallic \cite{Wolgast13,Kim13}. In ARPES, surface states were found at the \Gbar\ and \Xbar\ points \cite{Xu13,Jiang13}, as well as in quantum interference data \cite{LiG}. ARPES data were consistent in the observation of a Fermi surface contour at \Xbar\ \cite{Xu13,Jiang13,Neupane13,Frantzeskakis13,Denlinger13,Min14}. The dispersion of a Dirac cone was, however, not observed at \Gbar\ or at \Xbar. 

Hlawenka {\it et al}. \cite{Hlawenka18} demonstrated by core-level spectra two chemically pure terminations, consistent with earlier work \cite{Denlinger13}. The B and Sm terminations are also observed in scanning tunneling microscopy and differences in the assignment of scanning tunneling microscopy topographies in the literature have been explained by an element sensitivity due to the bias voltage \cite{Herrmann20}. Hlawenka {\it et al}. observed that the \Gbar\ state has a massive dispersion for Sm termination and is sensitive to contamination indicating that it is topologically trivial. For the B termination, it shows a Rashba splitting which is incompatible with a topological state. Since one \Gbar\ and two \Xbar\ Dirac cones have been predicted at the surface \cite{Alexandrov13,Takimoto11,LuF13}, it is actually unimportant for the topological assignment whether the contour at \Xbar\ is trivial. 

The closing of the 5$d$-4$f$ hybridization gap with increasing temperature has been observed in ARPES \cite{Denlinger13,Min14,Hlawenka18}. It was found that the surface state at \Xbar\ is a surface shifted hybridized 5$d$-4$f$ dispersion, which is shifted by 10 meV to higher binding energy for B termination and by about the same amount to lower binding energy for Sm termination \cite{Hlawenka18}. These shifts, which are consistently observed by scanning tunneling spectroscopy \cite{Herrmann20,Matt20}, are enough to bridge the bulk band gap and explain the robust surface metallicity despite the trivial character of the surface state at \Xbar. 

The results by Hlawenka et al. do not explain why Xu et al. \cite{XuN14} observe a spin texture expected for a TSS  for the \Xbar\ contour. A similar spin texture was reported for the non-polar (111) surface \cite{Ohtsubo19}. 
Ohtsubo et al. \cite{Ohtsubo19} noted that the measurement temperatures of 15-20 K \cite{XuN14,Ohtsubo19} would conflict with the gap energy scale by Hlawenka et al. if confirmed. A spin polarization has also been reported for YbB$_{12}$ as indication for a TSS   \cite{Hagiwara16}. 

Another open question in SmB$_6$, concerns the umklapp features seen in ARPES. They are consistent with the $(2\times1)$ reconstruction of the Sm terminated surface but appear also on the B terminated surface, which shows only a $(1\times1)$ topography \cite{Herrmann20}. 
In view of the trivial character of the observed \Gbar\ state, some researchers propose that the Fermi surface contour around \Gbar, assigned so far to umklapp from the \Xbar\ contour \cite{Xu13,Jiang13,Hlawenka18}, is instead the missing Dirac cone to render SmB$_6$ topological 
\cite{Ohtsubo19,Thunstroem21,LiLu20}. 

Recently, FeSb$_2$ has been proposed as another candidate for a topological Kondo insulator \cite{XuKJ20}. Here the Fe is mixed valent and the hybridization gap occurs with its 3d states. The system also shows a conductivity plateau similarly to SmB$_6$. Other suggested compounds include Fe$_2$VAl, FeSi, and FeGe$_3$ \cite{XuKJ20,Tomczak18}. 

\subsection{{\normalfont\normalsize\bfseries XI. Topological superconductors}} 


ARPES has been playing an important role in the investigation of superconductors, in particular at high energy resolution to access the superconducting gap and determine the superconducting pairing symmetry, the pseudogap above the critical temperature and other properties. 

In a topological superconductor the superconducting gap plays the role of the bulk gap in the TI {\braun and the topological superconductor can be
viewed as a TI  with particle-hole symmetry} \cite{Roy08,Schnyder08,QiRMP11}. The simplest scenario in which they can be expected to emerge is that where spin-polarised topological surface or edge states span across the Fermi level of a superconducting material (with the superconducting phase possessed either intrinsically or proximity induced). An overall anti-symmetric pairing is then possible between electrons populating the TSS branches at $\pm k$. Alternatively, Rashba surface states in magnetic systems could produce similar results~\cite{elliot_colloquium_2015, sato_topological_2017}. Indeed, time-reversal breaking topological superconductors could provide a platform for topological quantum computing making use of their non-Abelian statistics \cite{Nayak08}. Proximity between a magnetic topological insulator and an s-wave superconductor has, e.g., been suggested for a topological quantum bit \cite{LianPNAS18}.

Intrinsic superconductors with $E_{\text{F}}$-pinned topological states at the Fermi level seem to be relatively sparse in nature. Of those that do exist, it is unlikely that an intrinsic superconducting phase can be present without (semi-)metallic ground states, thus precluding the use of genuine TIs for this purpose. Despite this, there are
{\braun several candidate topological superconductors currently under investigation by ARPES.} 
For the Fe-based superconductor Fe(Te,Se), a bulk band inversion involving strong spin-orbit interaction
was predicted \cite{WangZ15,XuG16}. This band inversion {\braun should lead} to Dirac cone topological surface states at \Gbar\ of FeTe$_{0.55}$Se$_{0.45}$\,(100). 
By high-resolution laser ARPES, this surface state was recently {\braun   reported} \cite{ZhangPScience18,ZhangPNP19}. By superconducting proximity to the bulk, the Dirac cone is expected to open a superconducting gap. An isotropic superconducting gap was measured in the upper Dirac cone at $T=2.4$ K \cite{ZhangPScience18}. 
{\braun It was recently argued, however, that FeTe$_{0.55}$Se$_{0.45}$ is a superconducting 3D Dirac semimetal hosting type-I and type-II Dirac points and that its electronic structure remains topologically trivial \cite{BorisenkoNPJQM20}.
Further increase in Te content would possibly create a band inversion, however, only with sub-meV sized gap \cite{BorisenkoNPJQM20}.} Similarly, the superconducting Dirac semimetal 1T-PdTe$_2$ has topological surface states that cross the Fermi level, but scanning tunneling spectroscopy/microscopy and muon spin relaxation experiments have found only conventional superconducting pairing to date~\cite{clark_fermiology_2018, leng_typeI_2019}.

\subsection{{\normalfont\normalsize\bfseries {\braunb XII. Ultrafast experiments}}}

Time-resolved pump-probe ARPES experiments of topological matter allow to study unoccupied states, in particular TSSs, and the dynamics of excited charge carriers including their relaxation mechanisms \cite{Sob12,Wan12}, the spin texture of unoccpied states \cite{San16b,JozwiakNC16,Clark21}, and novel light-induced emergent phenomena \cite{WangScience13}.

Upon ultrafast excitation {\braunb of a TI} with infrared {\braunb ($\sim$1.5 eV)} laser pulses of $50$ fs, electrons from the Dirac cone as well as the bulk valence band electrons occupy within a few 100 fs states above the Fermi level of the Dirac cone as well as the bulk conduction band are excited. Within a few 100 fs they occupy states above the Fermi level of the Dirac cone as well as the bulk conduction band, as seen with probe pulses of $\approx$ 6 eV \cite{Sob12,Wan12}. For these experiments, conventional n-type Bi$_2$Se$_3$ has been used \cite{Wan12} as well as special p-type Bi$_2$Se$_3$ samples \cite{Sob12}, see Fig.~\ref{FigUltrafast}(a). Immediately after the excitation, a thermalized electron distribution is reached in the TSS and in the bulk conduction band \cite{Wan12}. The thermalization, which can be probed at the Fermi-Dirac cutoff of the excited electron distribution, occurs due to electron-electron scattering. Since the maximum population of the TSS of p-type Bi$_2$Se$_3$ was reached at $\sim700$ fs, higher-lying states are likely involved in the population \cite{Sob12}. The decay is noticeable after $\sim$2 ps. This timescale is identified with energy transfer from the electrons to the lattice and thus characteristic for electron-phonon scattering in the bulk of Bi$_2$Se$_3$. The dynamics persist, however, at a timescale $> 10$ ps. The reason is a second, slower component below the bulk conduction band edge. This slow component decays with the same timescale as the bulk conduction band population because the TSS is continuously filled by the slowly decaying bulk conduction band occupation \cite{Sob12}.

The combination of time- and spin-resolved ARPES allows to study the spin texture of unoccupied TSSs and the bulk conduction band. Again, $p$-type samples such as Sb$_2$Te$_3$ {\braunb leading to an unoccupied upper and largely occupied lower Dirac cone} \cite{ZhuSR15} are advantageous for studying the decay processes  \cite{San16b}. The spin texture forbids a vertical transition within the Dirac cone. For horizontal transitions involving phonons, the limitation is the maximum phonon energy, which is about 25 meV for Sb$_2$Te$_3$, similar to Bi$_2$Se$_3$ \cite{Hat11}. The deceleration of the recombination due to lack of available states is sometimes referred to as electron bottleneck effect. It can be seen directly in the data as enhanced transient photoemission intensity above the Dirac point. Measured relaxation rates were 5 to 10 ps 
 for Bi$_2$Se$_3$  \cite{Sob12,Wan12},  50 ps for p-type Bi$_{2.2}$Te$_3$   \cite{Haj14}, and  1.5 ps for  Sb$_2$Te$_3$ \cite{San16b}. Experiments with spin resolution allow to conclude whether spin manipulation with circularly polarized light, confirmed for continuous 6 eV ultraviolet light in Bi$_2$Se$_3$ \cite{JozwiakNatPhys13}, will extend to the infrared range. Spin- and time-resolved ARPES of Sb$_2$Te$_3$ shows that a circularly polarized pump pulse turns the spins of the Dirac cone from in plane to out of plane in the transient state \cite{San16b}, see Fig.~\ref{FigUltrafast}(c,e). Similar to the direct photoemission experiment at 6 eV \cite{JozwiakNatPhys13, SanchezPRX14} the spin orientation is determined by the light helicity (C$+$ and C$-$). This means that the spin manipulation does not only affect the final state of photoemitted electrons but rotates also the spin of electrons inside of the sample, an effect that could be employed for optospintronics \cite{San16b}. This has been confirmed   also for excitation into higher-lying surface states and resonances \cite{Soi19}. It was also possible to measure the {\braunb decay of the} spin polarization of the excited carriers which occurs twice as fast as the spin-averaged intensity, on a time scale of 0.4 ps as compared to 1 ps for the electron population \cite{San16b}. 
 The constant indirect filling of the bulk conduction band \cite{Sob12} probably  leads to a depolarization since it supplies unpolarized electrons from the bulk valence band \cite{San16b}.

 The above is an example for the importance of the coupling of surface and bulk states. The coupling of surface resonances to the bulk was found to be comparatively weak for Sb$_2$Te$_3$ \cite{Sei15} and Bi$_2$Se$_3$ \cite{Cac15}. For the n-doped systems Bi$_2$Se$_3$ and Bi$_2$Te$_3$, infrared light can excite electrons directly into higher lying surface resonances \cite{Cac15} which plays a crucial role in the spin dynamics \cite{Cac15, San17}. 
 
 When the stoichiometry is tuned to the bulk insulating state in
 (Bi,Sb)$_2$Te$_3$,   the   decay time increases dramatically from a few ps to several 100 ps   \cite{San17a,SumidaSR17}. A persistent photovoltage is observed and interpreted as caused by band bending \cite{San17a}. Thermalized electrons cannot diffuse from the surface state into the bulk while bulk holes diffuse to the surface. This special situation due to the intrinsically metallic surface has been suggested for use in photovoltaics \cite{San17a}.
 However, when Sb$_2$Te$_3$ is magnetically doped by V, the decay time {\braunb decreases again} strongly due to additional scattering paths, e. g., {\braunb at 0.33 eV} above the Fermi level from $\sim$3 ps to $\sim$ 100 fs \cite{SumidaNJP19}

In order to make use of topological insulators in spintronics, it is necessary to generate spin currents or spin-polarized currents, for example for spin-transfer torques as generated in TI/ferromagnet interfaces \cite{Mel14}. It has been reported that spin polarized currents can be triggered and controlled by circularly polarized infrared radiation of 1.5 eV  \cite{McI12} but this result {\braunb is awaiting confirmation by spectroscopic} methods. The intensity enhancement in ARPES upon ultrafast excitation  \cite{Sob12,Wan12}  becomes visibly asymmetric {\braunb in electron momentum {\bf k}$_\parallel$} for Sb$_2$Te$_3$ when the infrared light pulse is circularly polarized  \cite{San16b}, see Fig.~\ref{FigUltrafast}(c,d). Because of the locking of spin and linear momentum in the Dirac cone \cite{Pau12},  this finding was interpreted as confirmation that an ultrafast spin-polarized current is excited by the circularly polarized infrared light \cite{San16b}.  With smaller excitation energies of 0.25 to 0.37 eV, the asymmetry in momentum is also strong but does not reverse by switching the light helicity \cite{Kur16}. On the other hand, it has been pointed out for Bi$_2$Se$_3$ that the helicity-dependent asymmetry, that is, the circular dichroism, has at the Fermi energy a threefold asymmetry in the ($k_x$, $k_y$) momentum plane and, therefore, cannot lead to an overall spin-polarized current \cite{Ket18}, see Fig.~\ref{FigUltrafast}(f).  

Time-resolved ARPES has also been used to investigate carrier-wave-driven currents in the TSS \cite{ReimannNature18}. 
With subcycle time resolution, the acceleration by the carrier wave of a THz pulse of electrons in the TSS of Bi$_2$Te$_3$ is probed. The branches appear shifted in energy and momentum with respect to each other marking the shift of the Fermi circle in momentum space, see Fig.~\ref{FigUltrafast}(g-i). The effect reverses when the electric field as marked in  Fig.~\ref{FigUltrafast}(h) changes sign. It is concluded that large ballistic mean free paths result from spin-momentum locking \cite{ReimannNature18}.

Strong coupling of electronic states with photons can lead to Floquet-Bloch states which are periodic in momentum and energy \cite{FaisalPRA97}. This has been demonstrated for the TSS of Bi$_2$Se$_3$ using intense ultrashort mid-infrared pulses the energy of which is below the bulk band gap  \cite{WangScience13,MahmoodNP16}. It leads to the appearance of Dirac cone replicas above and below the original Dirac cone spaced by the driving photon energy, see Fig.~\ref{FigUltrafast}(b). Important for the confirmation as Floquet-Bloch states was the observation of hybridization gaps between these side bands \cite{WangScience13}. In addition, a gap at the Dirac point was created using circularly polarized light and assigned to time-reversal symmetry breaking and the creation of a photoinduced quantum anomalous Hall phase  \cite{WangScience13}.

\subsection{{\normalfont\normalsize\bfseries XIII. Outlook}} 

We end by highlighting 
a few aspects that we consider important for the future development of ARPES in topological materials.  

The theoretical prediction of topological materials based on their band structure has recently been performed  systematically \cite{BradlynTopQuantChemNature17,PoNC17,KruthoffPRX17}.
Under the theory of topological quantum chemistry, band structures and symmetry groups are systematically
analyzed by graph theory and group theory, respectively, and topologically nontrivial band structures can be systematically discovered. The nontrivial topology must be inherent in the band structure - as opposed to topological phases such as spin liquids - and electron correlation must be weak. ARPES will continue to directly verify these predictions and will contribute to explore the effects of electron correlation. 
 
Micro- and nano-ARPES (see the review by Hofmann \cite{HofmannReview21}) will lift the constraints to grow large single crystals. Small crystalline domains or small single crystals can be studied but the surface preparation by cleaving will become more challenging and novel approaches will be required. Micro/nano-ARPES will also enable to distinguish differently stacked local regions in stacked 2D materials such as TMDs, MXenes \cite{Gogotsi19}.

In the general framework of future electronics, one can say that the transistor-based electronics, which is currently in use, faces limitations when it comes to further downsizing. Additionally, its operation relies on relatively uncomplicated materials. In contrast, topological and other quantum materials possess  distinct sensitivity to external influences. In the case of topological protection, this sensitivity can be extremely low. On the other hand, the sensitivity can be significantly high.
In-operando ARPES can uncover such substantial state changes in response to electric current, bias voltage, electric field, temperature, and strain. In this way, topological phase transitions can be driven, and the effects may be leveraged in innovative devices. The phase space for the choice of the ideal material is enormously expanded in this way, and a material can be picked and further properties induced separately. 
This includes proximity effects as discussed above in the context of topological superconductivity. 
Dissipationless quantized edge states and functionalities required for topological quantum computations could thus be investigated in device-like structures. 

Hence, another use of the high spacial resolution of micro/nano-ARPES will relate to \textit{inoperando} use of ARPES \cite{HofmannReview21}. It has been shown for graphene and bilayer graphene stacked with h-BN that the effect of gating   can be resolved \cite{JouckenNL19,NguyenNature19,CurcioPRL20}.
It has also been achieved for the flat band of twisted bilayer graphene  \cite{JonesAdvMat20}
and for WSe$_2$ \cite{NguyenNature19}. 
 
It would be a strong advantage to determine the nontrivial topology directly in ARPES. 
For topological matter simulated by ultracold atomic gases, the quantization of circular dichroism in photoabsorption could be demonstrated
as distinct differentiation between topologically trivial and nontrivial phases \cite{AsteriaNP19}. 
Deducing the Berry phase from the CDAD has been proposed and calculated for 
graphene \cite{LiuChangPRL11,GierzNL12,SchuelerSA20}, monolayer WTe$_2$   \cite{MuechlerPRX16}
and  bulk 2H–WSe$_2$ \cite{ChoSPRL18}. 
It will be interesting to see how final-state effects can be treated and whether it will be possible to investigate arbitrary 3D materials.

\bibliographystyle{apsrev4-1}

\end{document}